\documentclass[]{aa}
\usepackage{times}
\usepackage{graphics}

\begin{document}

   \thesaurus{09          
              (02.03.3;   
               02.08.1;   
               02.12.1;   
               06.01.1;   
               06.07.2;   
               06.16.2)}   

   \title{Line formation in solar granulation}
   \subtitle{II. The photospheric Fe abundance}

   \author{M. Asplund\inst{1,2}, \AA. Nordlund\inst{3},
           R. Trampedach\inst{4}, and R.F. Stein\inst{4}
          }

   \offprints{Martin Asplund (martin@astro.uu.se)}

   \institute{
              NORDITA, 
              Blegdamsvej 17, 
              DK-2100 ~Copenhagen {\O}, 
              Denmark        
              \and
              present address: Uppsala Astronomical Observatory,
              Box 515,
              SE-751 20 ~Uppsala,
              Sweden
              \and
              Astronomical Observatory, NBIfAFG, 
              Juliane Maries Vej 30,
              DK-2100 ~Copenhagen \O, 
              Denmark
              \and
              Dept. of Physics and Astronomy, 
              Michigan State University, 
              East Lansing, MI 48823, USA
              }

   \date{Received: January 24, 2000; accepted: May 4, 2000 }

\authorrunning{M. Asplund et al.}
\titlerunning{Solar line formation: II. The photospheric Fe abundance}

   \maketitle

   \begin{abstract}

The solar photospheric Fe abundance has been determined using 
realistic ab initio 3D, time-dependent, hydrodynamical model atmospheres.
The study is based on the excellent agreement between the
predicted and observed line profiles directly rather than
equivalent width, since the 
intrinsic Doppler broadening from the convective motions and oscillations
provide the necessary non-thermal broadening.
Thus, three of the four hotly debated parameters (equivalent
widths, microturbulence and damping enhancement factors) in the center 
of the recent solar Fe abundance dispute regarding Fe\,{\sc i} lines 
no longer enter the analysis, leaving the transition probabilities as the
main uncertainty. Both Fe\,{\sc i} (using the samples of lines of both
the Oxford and Kiel studies) and Fe\,{\sc ii} lines have been
investigated, which give consistent results:
${\rm log}\, \epsilon_{\rm FeI}  = 7.44 \pm 0.05 $ and
${\rm log}\, \epsilon_{\rm FeII} = 7.45 \pm 0.10.$
Also the wings of strong Fe\,{\sc i} lines return consistent abundances,
${\rm log}\, \epsilon_{\rm FeII} = 7.42 \pm 0.03,$ but due to the
uncertainties inherent in analyses of strong lines we give this
determination lower weight than the results from 
weak and intermediate strong lines.
In view of the recent slight downward revision of the meteoritic
Fe abundance
${\rm log}\, \epsilon_{\rm Fe} = 7.46 \pm 0.01$, the agreement between
the meteoritic and photospheric values is very good,
thus appearingly settling the debate over the photospheric Fe abundance
from Fe\,{\sc i} lines.

   \end{abstract}

\section{Introduction}

The solar iron abundance is of fundamental importance as it
provides the standard to which all other elemental abundances
in stars are compared. Furthermore, since Fe is the dominating
contributor to the total line-blanketing and a significant 
electron donor for late-type stars such as the Sun, the exact
value of the Fe abundance influences the overall photospheric structure.
Thus Fe indirectly affects the emergent spectrum and the derived 
abundances for other elements as well. In spite of its significance
and after many previous investigations, the solar Fe content 
is still, astonishingly enough, debated on the level of 0.2\,dex
(Blackwell et al. 1995a,b; Holweger et al. 1995). Not even the
basic reasons for this large discrepancy have been properly
understood, though it is commonly blamed on differences in
adopted parameters for the analysis ($gf$-values, equivalent
widths, collisional damping parameters, microturbulent velocities)
as well as subtle differences in computer codes (Kostik et al. 1996).
Such dissonance even for the Sun naturally rises
concern regarding derived stellar abundances with claimed accuracies of 
$0.05\,$dex or less. 

Recently, with the advent of accurate $gf$-values for weak
Fe\,{\sc ii} lines (Heise \& Kock 1990; Holweger et al. 1990; 
Bi{\'e}mont et al. 1991; Hannaford et al. 1992; 
Raassen \& Uylings 1998; Schnabel et al. 1999) and improved
treatment of the collisional damping of Fe\,{\sc i} lines
(Milford et al. 1994; Anstee et al. 1997),  
there seems to be some convergence towards finding consistency
between the photospheric and the meteoritic
Fe abundances (Grevesse \& Sauval 1998, 1999).
The studies by the Oxford group (Blackwell et al. 1995a,b and
references therein), however, stand out with their 
distinguished high value of log$\,\epsilon_{\rm Fe\,I}=7.64$ 
\footnote{On the customary logarithmic abundance scale defined
to have log$\,\epsilon_{\rm H}=12.00$}
rather than the current best estimate of 
log$\,\epsilon_{\rm Fe}=7.50$ for the meteoritic abundance 
(Grevesse \& Sauval 1998, but see Asplund 2000, hereafter Paper III)
as determined from carbonaceous chondrites of type 1 (C1 chondrites). 

It is sobering to remember that all of the above-mentioned investigations
rely on several approximations and assumptions not necessarily
justified in the case of the Sun. 
Traditional abundance analysis of stars are based on one-dimensional
(1D), theoretical model atmospheres constructed under the assumptions
of plane-parallel geometry (or spherical geometry for stars with
extended atmospheres), hydrostatic equilibrium (or steady state stellar
winds for hot stars), flux constancy 
and with the convective energy transport computed through the
mixing length theory (B\"ohm-Vitense 1958) 
or some close relative thereof (e.g. Canuto \& Mazzitelli 1991), with all
their limitations and free parameters. For late-type stars 
the simplifying assumption of LTE is also normally adopted 
(cf. Gustafsson \& J{\o}rgensen 1994 for a review of
stellar modelling of late-type stars). 
For the Sun, information of the emergent spectrum,
e.g. details of the limb-darkening, may be utilized to
construct a semi-empirical model atmosphere, such as the widely used 
Holweger-M\"uller (1974) model 
(the assumptions of 1D, plane-parallel
geometry, hydrostatic equilibrium and LTE have here, however, been retained).

A closer inspection of the solar photosphere reveals that none
of these assumptions are strictly correct: the solar surface
is dominated by the granulation pattern reflecting the convection
zone deeper inside, which results in an evolving inhomogeneous surface
structure with prominent velocity fields between warm upflows (granules)
and cool downflows (intergranular lanes) with very different 
temperature gradients
(e.g. Stein \& Nordlund 1998). Of course there are also
regions with significantly enhanced magnetic field strengths,
which may influence the emergent spectrum. It is therefore not
surprising that none of the available 1D model atmospheres, 
including the Holweger-M\"uller (1974) model, satisfactory predict
simultaneously all of the various observational diagnostics
(limb-darkening, flux distribution, H-lines etc)
even for the Sun 
(e.g. Blackwell et al. 1995a; Allende Prieto et al. 1998)
Naturally, different species are affected differently by the
granulation and its heterogeneous nature. Lines from the dominant
ionization stage (Fe\,{\sc ii} in the case of the Sun) will be
less influenced by the details of the atmospheric
structure while lines from other ionization stages are more
sensitive. Furthermore, lines from minority species will
in general be more susceptible to departures
from LTE, which can be expected to be more pronounced in inhomogeneous
atmospheres compared with 1D model atmospheres
(cf. discussion in Kostik et al. 1996). 

Furthermore, abundance analyses normally proceed with additional
assumptions when synthesizing the spectrum. In order to approximately
account for the photospheric velocity fields and produce the
needed extra line broadening, both
a microturbulent velocity $\xi_{\rm turb}$ --  
supposedly representing small-scale velocities -- as well
a macroturbulent velocity -- reflecting large-scale 
motions not present
in the model atmospheres -- are applied to the spectral synthesis.
The exact shapes of these additional broadening recipes to be 
convolved with the synthetic spectrum also remain poorly understood
(cf. Gray 1992).
Finally, the treatment of collisional line broadening normally stems from
the approach by Uns\"old (1955), enhanced by an
ad-hoc factor to account for the still lacking 
amount of broadening
for strong lines. With recent quantum mechanical calculations 
(Anstee \& O'Mara 1991, 1995; 
Barklem \& O'Mara 1997; Barklem et al. 1998) 
the introduction of the unknown damping
enhancement factor may no longer be necessary however, at least not for
lines of neutral species.

There is therefore no doubt that traditional abundance determination
are built on somewhat shaky grounds, which need to be verified
by more detailed calculations. With the recent progress in ab-initio
numerical multi-dimensional hydrodynamical simulations 
of surface convection of stars 
(e.g. Stein \& Nordlund 1989, 1998; Nordlund \& Dravins 1990; 
Atroshchenko \& Gadun 1994; Freytag et al. 1996; Kim \& Chan 1998; 
Asplund et al. 1999a; Ludwig et al. 1999; Trampedach et al. 1999)
there is fortunately an alternative to classical model atmospheres.
Such inhomogeneous atmospheres self-consistently calculate the
convective energy transport and the 
velocity and temperature structures, making the concepts of
mixing length parameters, 
microturbulent and macroturbulent velocities obsolete. 
The high-degree of realism of these convection simulations
is supported by the fact that they successfully reproduce the
solar granulation pattern and statistics (Stein \& Nordlund 1989, 1998), 
helioseismological constraints such as p-mode frequencies
and depth of the convection zone (Rosenthal et al. 1999), 
and spectroscopic diagnostics such as flux-distribution, limb-darkening
and detailed line profiles and asymmetries, even on an absolute
wavelength scale (Asplund et al. 1999b;
Asplund et al. 2000b, hereafter Paper I). In the present
paper we apply such granulation simulations of the 
Sun to the problem of the solar Fe abundance, utilizing both
weak lines and the wings of strong lines.
In order to minimize the impact of possible
departures from LTE both lines of neutral and ionized Fe
have been investigated. By fitting the line profiles the uncertainties
introduced with equivalent widths can be avoided. 
Furthermore, whenever possible the
improved collisional broadening treatment of Anstee \& O'Mara 
(1991) has been used.

\section{3D model atmospheres and spectral line calculations}

The procedure for calculating the spectral line transfer is the
same as in Paper I and therefore only a short summary will be given here.
For additional information on the details of the convection simulations
and the 3D spectral synthesis, the reader is referred to Paper I. 

Realistic ab-initio numerical hydrodynamical simulations of the solar surface
convection have been performed and used as 3D, time-dependent, inhomogeneous 
model atmospheres with a self-consistent description of the convective
flow and temperature structure in the photosphere. 
A state-of-the-art equation-of-state (Mihalas et al. 1988) has been
used together with the 3D equation of radiative transfer which included
the effects of line-blanketing (Nordlund 1982) with up-to-date
continuous (Gustafsson et al. 1975 with subsequent
updates) and line opacities (Kurucz 1993).
The original simulation has a resolution of 200\,x\,200\,x\,82,
which was interpolated to a grid with dimension 50\,x\,50\,x\,82
to ease the computational burden in the spectral line calculations.
Simultaneously the vertical resolution was improved by only extending
down to depths of about 700\,km compared with the initial 2.9\,Mm.
Various test ensured that this procedure had no effect on the 
resulting profiles. The convection simulation used for the 
spectral synthesis here and in Paper I covered about 50\,min
on the Sun. For the present purposes the time coverage is sufficient to 
obtain properly spatially and temporally averaged line profiles,
as verified by test calculations; even intervals
as short as 10\,min result in abundances within 0.02\,dex of
the estimates using the whole time-sequence. 
The resulting effective temperature is very close to
the nominal solar value, $T_{\rm eff} = 5767 \pm 21$\,K, while
the adopted surface gravity was log\,$g = 4.437$ [cgs]. For the
computation of background continuous opacities and equation-of-state,
a standard solar chemical composition was used (Grevesse \& Sauval 1998).
In particular the assumed He abundance (10.93) was consistent with
the helioseismological evidence (e.g Basu 1998; Grevesse \& Sauval 1998) 
though the exact
value is of no practical importance for the present investigation.

In the present investigation only intensity spectra at 
solar disk center ($\mu = 1.0$) were considered,
which have been calculated for every column of the snapshots, before 
spatial and temporal averaging and normalization.
The assumption
of LTE in the ionization and excitation balances and for the
source function ($S_\nu = B_\nu$)
have been made throughout in the line transfer calculations. 
The line profiles were computed for 141 velocities around the 
laboratory wavelength with a interval of 0.2 (weak and intermediate strong
lines) or 1.5-2.0\,km\,s$^{-1}$ (strong lines); one additional point was 
computed without consideration of the line to estimate the continuum intensity
necessary for the normalization.
All lines were calculated with three different abundances 
(${\rm log}\, \epsilon_{\rm Fe} = 7.30$, 7.50 and
7.70) from which the final profile with the 
correct line strength was interpolated from a $\chi^2$-analysis of the whole
profile in a similar fashion to the study of Nissen et al. (2000); 
test calculations ensured that 
the abundance step was sufficiently small not to introduce any 
significant errors in the derived abundances 
($\Delta {\rm log}\, \epsilon_{\rm Fe } \ll 0.005$\,dex).

\section{Line data and observed solar spectrum}

The accuracy of the final results naturally depend not only on the
degree of realism of the model atmosphere but also on the quality of
the necessary input atomic data.
The choice of transition probabilities
for the lines is a delicate matter, with many recent discussions
of pros and cons in the literature. The $gf$-values for 
Fe\,{\sc i} lines of the Oxford group (Blackwell et al. 1995a and
references therein), the Hannover-Kiel workers (Holweger et al. 1995
and references therein) and O'Brian et al. (1991)
are of excellent internal consistency and all agree to within
0.03\,dex on average (Lambert et al. 1996).
The last source has larger quoted uncertainties in general,
which is also evident in the significantly larger 
scatter in the derived abundances. 
For completeness we have included both remaining samples in the analysis;
the overlap of lines with quoted equivalent widths $W_\lambda \le 10$\,pm
is limited to nine lines, which differ by 0.028\,dex on average.
However, given the in general good reputation of
furnace measurements and the fact that concern
regarding the quality of the $gf$-values from the Hannover-Kiel
group recently has been voiced (Kostik et al. 1996; Anstee et al. 1997), 
we tend to give
the results obtained with the Oxford transition probabilities 
greater weight. Also the scatter in the derived abundances 
is slightly larger when adopting the Hannover-Kiel $gf$-values. 
When deriving the Fe abundance from the wings of strong Fe\,{\sc i}
lines, we have been guided by the quality measures quoted 
by Anstee et al. (1997) and selected the most suitable lines 
(in total 14 lines).
The $gf$-values for these lines are taken from Blackwell et al. (1995a)  
and O'Brian et al. (1991).

\begin{table*}[t!]
\caption{The adopted line data and individually derived abundances 
for the weak and intermediate strong Fe\,{\sc i} lines
\label{t:fei}
}
\begin{tabular}{lcccccccc} 
 \hline
Wavelength$^{\rm a}$ & $\chi_{\rm l}$$^{\rm a}$ & log\,$gf$ & ref. & 
log\,$\gamma_{\rm rad}$$^{\rm a}$ &
lower & upper & $W_\lambda^{\rm b}$ & log\,$\epsilon_{\rm Fe}$ \\
$$ [nm]    & [eV]   &        &  gf$^{\rm b}$ &        & level$^{\rm a}$ 
& level$^{\rm a}$ & [pm] $$ & \\
\hline 
 438.92451 &  0.052 & -4.583 & B &  4.529 & s & p &   7.17 &  7.43 \\
 444.54717 &  0.087 & -5.441 & B &  4.529 & s & p &   3.88 &  7.42 \\
 524.70503 &  0.087 & -4.946 & B &  3.894 & s & p &   6.58 &  7.42 \\
 525.02090 &  0.121 & -4.938 & B &  3.643 & s & p &   6.49 &  7.45 \\
 570.15444 &  2.559 & -2.216 & B &  8.167 & s & p &   8.51 &  7.53 \\
 595.66943 &  0.859 & -4.605 & B &  4.433 & s & p &   5.08 &  7.43 \\
 608.27104 &  2.223 & -3.573 & B &  6.886 & s & p &   3.40 &  7.42 \\
 613.69946 &  2.198 & -2.950 & B &  8.217 & s & p &   6.38 &  7.46 \\
 615.16182 &  2.176 & -3.299 & B &  8.190 & s & p &   4.82 &  7.42 \\
 617.33354 &  2.223 & -2.880 & B &  8.223 & s & p &   6.74 &  7.44 \\
 620.03130 &  2.608 & -2.437 & B &  8.013 & s & p &   7.56 &  7.49 \\
 621.92808 &  2.198 & -2.433 & B &  8.190 & s & p &   9.15 &  7.45 \\
 626.51338 &  2.176 & -2.550 & B &  8.220 & s & p &   8.68 &  7.45 \\
 628.06182 &  0.859 & -4.387 & B &  4.622 & s & p &   6.24 &  7.46 \\
 629.77930 &  2.223 & -2.740 & B &  8.190 & s & p &   7.53 &  7.44 \\
 632.26855 &  2.588 & -2.426 & B &  8.009 & s & p &   7.92 &  7.51 \\
 648.18701 &  2.279 & -2.984 & B &  8.190 & s & p &   6.42 &  7.47 \\
 649.89390 &  0.958 & -4.699 & B &  4.638 & s & p &   4.43 &  7.43 \\
 657.42285 &  0.990 & -5.004 & B &  4.529 & s & p &   2.65 &  7.38 \\
 659.38706 &  2.433 & -2.422 & B &  7.936 & s & p &   8.64 &  7.53 \\
 660.91104 &  2.559 & -2.692 & B &  7.905 & s & p &   6.55 &  7.49 \\
 662.50220 &  1.011 & -5.336 & B &  4.403 & s & p &   1.36 &  7.36 \\
 675.01523 &  2.424 & -2.621 & B &  6.886 & s & p &   7.58 &  7.48 \\
 694.52051 &  2.424 & -2.482 & B &  7.196 & s & p &   8.38 &  7.48 \\
 697.88516 &  2.484 & -2.500 & B &  6.886 & s & p &   8.01 &  7.49 \\
 772.32080 &  2.279 & -3.617 & B &  6.848 & s & p &   3.85 &  7.55 \\
 504.42114 &  2.851 & -2.059 & H &  8.009 & p & s &   7.50 &  7.45 \\
 525.34619 &  3.283 & -1.573 & H &  7.875 & p & s &   8.10 &  7.40 \\
 532.99893 &  4.076 & -1.189 & H &  7.659 & d & p &   5.60 &  7.47 \\
 541.27856 &  4.434 & -1.716 & H &  8.226 & p & d &   1.78 &  7.45 \\
 549.18315 &  4.186 & -2.188 & H &  8.158 & d & p &   1.06 &  7.41 \\
 552.55444 &  4.230 & -1.084 & H &  8.382 & p & s &   5.80 &  7.39 \\
 566.13457 &  4.284 & -1.756 & H &  7.908 & p & s &   1.98 &  7.38 \\
 570.15444 &  2.559 & -2.130 & H &  8.167 & s & p &   8.60 &  7.45 \\
 570.54648 &  4.301 & -1.355 & H &  8.290 & p & s &   3.90 &  7.38 \\
 577.84531 &  2.588 & -3.440 & H &  8.167 & s & p &   1.95 &  7.36 \\
 578.46582 &  3.396 & -2.530 & H &  7.877 & p & s &   2.50 &  7.39 \\
 585.50767 &  4.607 & -1.478 & H &  8.281 & p & d &   2.10 &  7.41 \\
 608.27104 &  2.223 & -3.590 & H &  6.886 & s & p &   2.82 &  7.44 \\
 615.16182 &  2.176 & -3.270 & H &  8.190 & s & p &   4.56 &  7.39 \\
 621.92808 &  2.198 & -2.422 & H &  8.190 & s & p &   8.70 &  7.44 \\
 624.06460 &  2.223 & -3.230 & H &  7.196 & s & p &   4.38 &  7.36 \\
 627.12788 &  3.332 & -2.703 & H &  8.074 & p & s &   2.09 &  7.40 \\
 629.77930 &  2.223 & -2.727 & H &  8.190 & s & p &   7.30 &  7.42 \\
 648.18701 &  2.279 & -2.960 & H &  8.190 & s & p &   6.30 &  7.45 \\
 658.12100 &  1.485 & -4.680 & H &  7.193 & s & p &   1.41 &  7.39 \\
 666.77114 &  4.584 & -2.112 & H &  8.158 & s & p &   0.89 &  7.55 \\
 669.91416 &  4.593 & -2.101 & H &  8.158 & s & p &   0.73 &  7.45 \\
 673.95220 &  1.557 & -4.790 & H &  7.176 & s & p &   1.03 &  7.30 \\
 675.01523 &  2.424 & -2.610 & H &  6.886 & s & p &   7.70 &  7.47 \\
 679.32593 &  4.076 & -2.326 & H &  7.622 & d & p &   1.10 &  7.39 \\
 680.42715 &  4.584 & -1.813 & H &  7.719 & s & p &   1.40 &  7.46 \\
 683.70059 &  4.593 & -1.687 & H &  7.719 & s & p &   1.54 &  7.44 \\
 685.48228 &  4.593 & -1.926 & H &  7.659 & s & p &   1.00 &  7.52 \\
 694.52051 &  2.424 & -2.440 & H &  7.196 & s & p &   8.20 &  7.44 \\
 697.19330 &  3.018 & -3.340 & H &  8.161 & s & p &   1.20 &  7.36 \\
 697.88516 &  2.484 & -2.480 & H &  6.886 & s & p &   7.90 &  7.47 \\
 718.91510 &  3.071 & -2.771 & H &  8.161 & s & p &   3.80 &  7.53 \\
 740.16851 &  4.186 & -1.599 & H &  7.847 & d & p &   4.10 &  7.50 \\
\hline 
\end{tabular}
\begin{list}{}{}
\item[$^{\rm a}$] From Nave et al. (1994) and 
the VALD data base (Kupka et al. 1999)
\item[$^{\rm b}$] From Blackwell et al. (1995a) (ref. gf=B) and Holweger et al. (1995) (ref. gf=H).
Note that $W_\lambda$ is only 
listed to allow indentification in Fig. \ref{f:fei} and is not used
for deriving abundances
\end{list}

\end{table*}

\begin{table*}[t!]
\caption{The adopted line data for the strong Fe\,{\sc i} lines
\label{t:feistrong}
}
\begin{tabular}{lccccccccc} 
 \hline
Wavelength$^{\rm a}$ & $\chi_{\rm l}$$^{\rm a}$ & log\,$gf$ & ref. &
log\,$\gamma_{\rm rad}$$^{\rm a}$ &
lower & upper & $\sigma^{\rm c}$ & $\alpha^{\rm c}$ & log\,$\epsilon_{\rm Fe}$ \\
$$ [nm]    & [eV]   &        & gf$^{\rm b}$ &        & level  & level     \\
\hline  
 407.17380 &  1.608 & -0.022 & B & 8.009 & s & p &   328 &  0.252 &  7.40$^{\rm d}$ \\
 438.35449 &  1.485 &  0.200 & B &  7.936 & s & p &   295 &  0.265 &  7.43$^{\rm d}$ \\
 441.51226 &  1.608 & -0.615 & B &  7.986 & s & p &   305 &  0.261 &  7.46$^{\rm d}$ \\
 489.07549 &  2.875 & -0.390 & O &  8.004 & p & s &   747 &  0.236 &  7.45 \\
 489.14922 &  2.851 & -0.110 & O &  8.009 & p & s &   739 &  0.236 &  7.45 \\
 491.89941 &  2.865 & -0.340 & O &  8.009 & p & s &   739 &  0.237 &  7.46$^{\rm d}$ \\
 495.72988 &  2.851 & -0.410 & O &  8.009 & p & s &   727 &  0.238 &  7.43 \\
 495.75967 &  2.808 &  0.230 & O &  8.009 & p & s &   713 &  0.238 &  7.43 \\
 523.29404 &  2.940 & -0.060 & O &  8.009 & p & s &   712 &  0.238 &  7.41 \\
 526.95376 &  0.859 & -1.321 & B &  7.185 & s & p &   237 &  0.249 &  7.38 \\
 532.80386 &  0.915 & -1.466 & B &  7.161 & s & p &   239 &  0.248 &  7.40 \\
 532.85317 &  1.557 & -1.850 & O &  6.848 & s & p &   282 &  0.252 &  7.40$^{\rm d}$ \\
 537.14897 &  0.958 & -1.645 & B &  7.152 & s & p &   240 &  0.248 &  7.41$^{\rm d}$ \\
 544.69170 &  0.990 & -1.910 & O &  7.152 & s & p &   241 &  0.248 &  7.40$^{\rm d}$ \\
\hline 
\end{tabular}
\begin{list}{}{}
\item[$^{\rm a}$] From Nave et al. (1994) and
the VALD data base (Kupka et al. 1999)
\item[$^{\rm b}$] From Blackwell et al. (1995a) (ref. gf=B) and O'Brian et al. (1991) (ref. gf=O)
\item[$^{\rm c}$] Collisional broadening data 
from Barklem (1999, private communication)
\item[$^{\rm d}$] Lines which are given half weight in the final abundance
estimate due to uncertainties introduced by blending lines, continuum placement,
radiation broadening and poorly developed damping wings
\end{list}

\end{table*}

\begin{table}[t!]
\caption{The adopted line data for the Fe\,{\sc ii} lines
\label{t:feii}
}
\begin{tabular}{lccccc} 
 \hline
Wavelength$^{\rm a}$ & $\chi_{\rm l}$$^{\rm a}$ & log\,$gf$$^{\rm b}$ & 
log\,$\gamma_{\rm rad}$$^{\rm a}$ &
$W_\lambda$$^{\rm b}$ & log\,$\epsilon_{\rm Fe}$ \\
$$ [nm]    & [eV]   &        &        & [pm]  &       \\
 \hline
 457.63334 &  2.844 & -2.94 &  8.612 &  6.80 &  7.42 \\
 462.05129 &  2.828 & -3.21 &  8.615 &  5.40 &  7.35 \\
 465.69762 &  2.891 & -3.59 &  8.612 &  3.80 &  7.40 \\
 523.46243 &  3.221 & -2.23 &  8.487 &  8.92 &  7.49 \\
 526.48042 &  3.230 & -3.25 &  8.614 &  4.74 &  7.63 \\
 541.40717 &  3.221 & -3.50 &  8.615 &  2.76 &  7.38 \\
 552.51168 &  3.267 & -3.95 &  8.615 &  1.27 &  7.35 \\
 562.74892 &  3.387 & -4.10 &  8.487 &  0.86 &  7.49 \\
 643.26757 &  2.891 & -3.50 &  8.462 &  4.34 &  7.38 \\
 651.60716 &  2.891 & -3.38 &  8.464 &  5.75 &  7.52 \\
 722.23923 &  3.889 & -3.36 &  8.617 &  2.00 &  7.60 \\
 722.44790 &  3.889 & -3.28 &  8.617 &  2.07 &  7.55 \\
 744.93305 &  3.889 & -3.09 &  8.612 &  1.95 &  7.28 \\
 751.58309 &  3.903 & -3.44 &  8.612 &  1.49 &  7.49 \\
 771.17205 &  3.903 & -2.47 &  8.615 &  5.06 &  7.41 \\
\hline 
\end{tabular}
\begin{list}{}{}
\item[$^{\rm a}$] From Johansson (1998, private communication) and 
the VALD data base (Kupka et al. 1999)
\item[$^{\rm b}$] From Hannaford et al. 1992. Note that $W_\lambda$ is only
listed here to allow easy identification in Fig. \ref{f:feii} and is
not used for the abundance determinations
\end{list}

\end{table}

For the Fe\,{\sc ii}
lines there are five recent sources for $gf$-values:
Heise \& Kock (1990), Hannaford et al. (1992), Bi{\'e}mont et al. (1991),
Raassen \& Uylings (1998) and Schnabel et al. (1999)
of which the first two and the last are 
based on experimental data while the remaining
two have been obtained from semi-empirical calculations. 
Again, the variations between the different compilations are relatively
small on average, though occasionally large on a line-by-line
comparison.
We tend to view the theoretical calculations with
some balanced scepticism due to the noticably larger scatter
in derived abundances when selecting the $gf$-values by
Bi{\'e}mont et al.; for the
nine lines in common between all five sources, the standard deviation
increases from 0.07\,dex for the values by Hannaford et al. and
Heise \& Kock to 0.13\,dex when using Bi{\'e}mont et al.'s predictions
(cf. also discussions in Hannaford et al. 1992 and Bell et al. 1994). 
Furthermore, with the data from Bi{\'e}mont et al. the derived 
Fe abundances show a distinct trend with wavelength, suggesting a
problem in the calculations.
With the more recent calculations by Raassen \& Uylings the scatter
is improved to a comparable level to the measured $gf$-values, but
the absolute scale is clearly offset ($-0.10$\,dex relative to
Hannaford et al.) compared with the other four
sources, and therefore we are hesitant to adopt these calculations here. 
An investigation of the reason for these differences seems worthwhile
(cf. Grevesse \& Sauval 1999).
The final choice between the remaining three 
compilations is somewhat arbitrary, but we have opted for
the measurements by Hannaford et al. (1992) as it includes two
additional lines (15 lines in total) 
and the absolute scale of their gf-values is in between the other two
sources; adopting instead the values by Heise \& Kock (1990) and 
Schnabel et al. (1999) 
would change the derived abundance by +0.04\,dex and -0.02\,dex,
respectively, but leave the line scatter essentially unaltered.
We note that the more recent lifetime measurements of 
Schnabel et al. have slightly smaller claimed uncertainties than in
Hannaford et al.. However, since the independent experiments of 
Guo et al. (1992) support the measured lifetimes of Hannaford et al.,
we will here retain the Hannaford et al. $gf$-values but keep in mind
that the derived Fe\,{\sc ii} abundances may be overestimated with 0.02\,dex
on average.  

For the collisional broadening from
hydrogen atoms the quantum mechanical calculations developed
by Anstee \& O'Mara (1991) have been applied for the Fe\,{\sc i} lines
for transitions between levels of type s-p, p-s, p-d, d-p, d-f, and f-d. 
The broadening cross-sections and their dependence on 
temperature have been kindly provided by Barklem (1999,
private communication) from
line-by-line calculations
(the data has subsequently been incorporated into the VALD
database, Barklem et al. 2000). For a few lines not individually computed,
the necessary data was obtained from interpolation in tables
provided by Anstee \& O'Mara (1995), Barklem \& O'Mara (1997) and
Barklem et al. (1998). 
The contribution from collisions with helium atoms have been
included by assuming that the cross-sections scale with the
polarizability of the perturbing atom in the same way as in the
van der Waal's theory; due to the lower abundance and velocities
of He atoms this contribution is, however, very small and does not
influence the calculated profiles. 
Since the theory has not yet been fully extended to 
transitions from ionized species we have to rely on the
normal van der Waal's broadening approximation by Uns\"old (1955)
with an additional enhancement factor $E=2.0$ for the Fe\,{\sc ii}
lines, which is typical to those adopted in earlier investigations
of solar Fe\,{\sc ii} lines 
(Holweger et al. 1990; Bi{\'e}mont et al. 1991; Hannaford et al. 1992).
Fortunately, since the Fe\,{\sc ii} lines are all weak,
the impact of different choices of E is minor: adopting
E=1.5 instead leads to only a 0.01\,dex increase in the mean abundance. 
Radiative damping was included either with values obtained from
VALD (Kupka et al. 1999) or calculated from
the classical formula using the $gf$-value of the transition; only in 
a few cases does the exact choice of the radiative damping influence
the results by more than 0.01\,dex. Stark broadening was not considered.

A summary of the adopted line data for the 
Fe\,{\sc i} and Fe\,{\sc ii} lines is found in 
Tables \ref{t:fei}, \ref{t:feistrong} and \ref{t:feii}. 
The central wavelengths for the Fe\,{\sc i} and Fe\,{\sc ii} lines were
taken from Nave et al. (1994) and Johansson 
(1998, private communication). When deriving elemental abundances
this choice is of course of minor importance, though it is crucial to
have accurate estimates when studying line asymmetries (Paper I).

One of the novel features of the current analysis is that the Fe
abundances are derived from a fit of the line profiles rather than
from equivalent widths as customary done. This is facilitated by
the excellent agreement between observed and predicted line shapes,
including the departures from perfect symmetry when including
the effects of Doppler shifts due to the convective flows (Paper I). 
No microturbulent or macroturbulent velocities therefore enter the spectral
synthesis, since the self-consistent velocity field of the simulation
is taken into account properly. 
Thereby we have managed to remove three of the four hotly debated parameters
($W_\lambda$, $\xi_{\rm turb}$ and E), which have been blamed for the
discordance in Fe abundance between the Oxford and Hannover-Kiel results.
For illustrative purposes (e.g. Fig. \ref{f:fei} and \ref{f:feii}) 
and for the selection of lines only, we have used the quoted equivalent
widths from the appropriate sources in the literature. We emphasize 
that they are not used when determining the Fe abundances.

For the comparison of line shapes the solar FTS disk-center intensity
atlas by Brault \& Neckel (1987) and Neckel (1999) 
has been used due to its superior
quality over the older Liege atlas by Delbouille et al. (1973) 
in terms of wavelength calibration 
(Allende Prieto \& Garc\'{\i}a L{\'o}pez 1998a,b).
In a few cases the continuum level was renormalized to better trace
the local continuum around the lines.

\section{Abundance from weak and intermediate strong Fe\,{\sc i} lines
\label{s:fei}}

\begin{figure*}[t!]
\resizebox{\hsize}{!}{\includegraphics{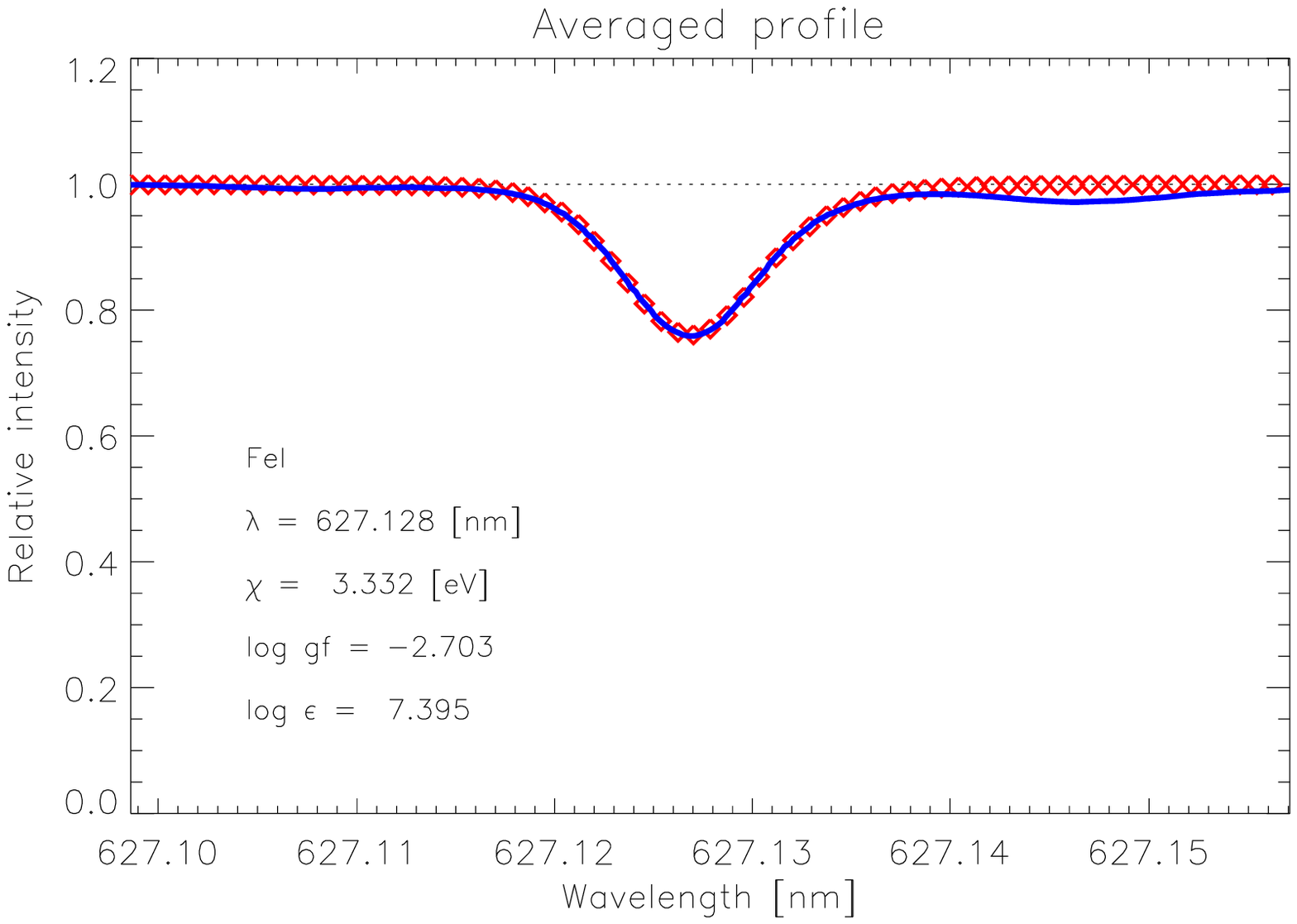}
\includegraphics{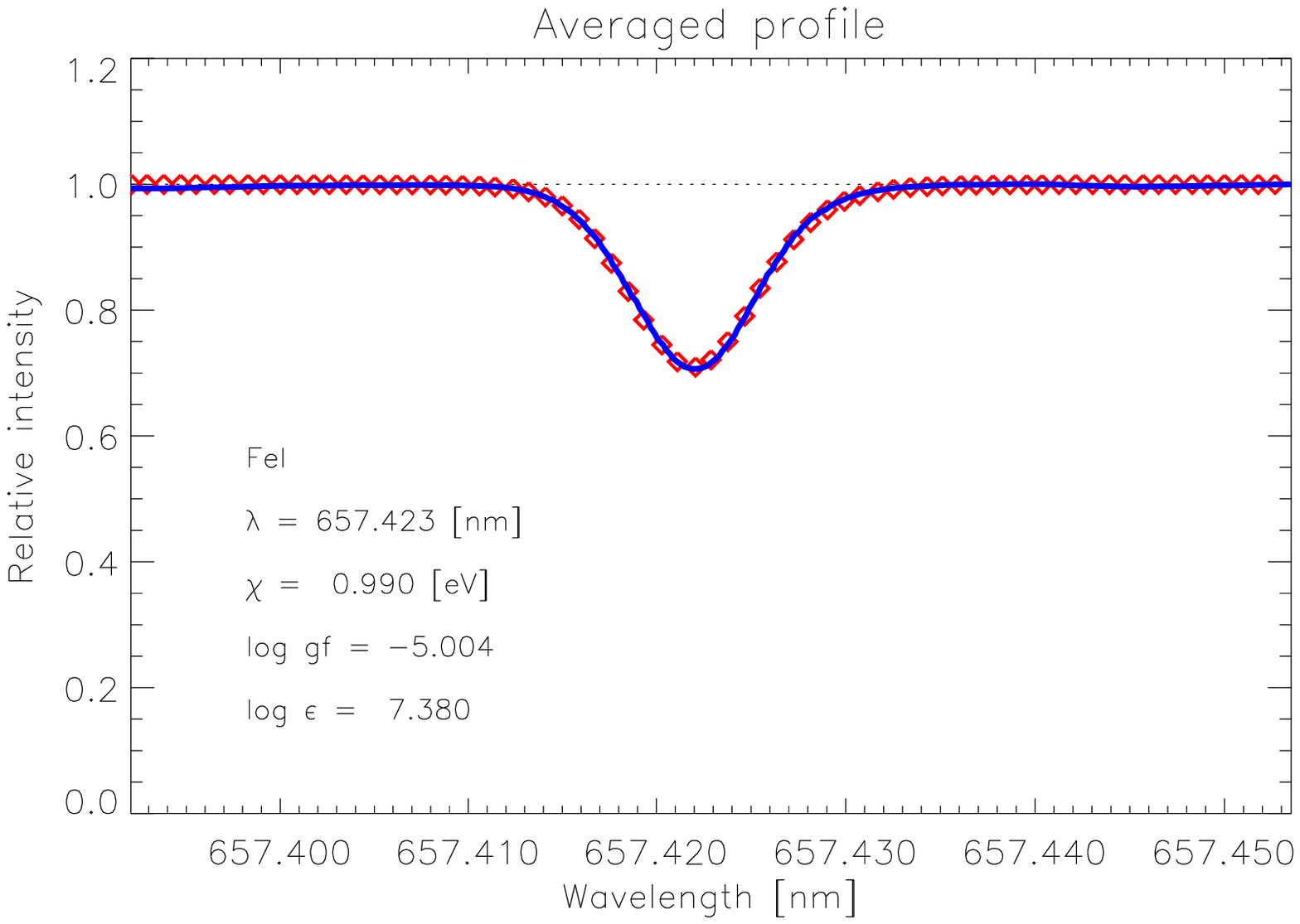}}
\resizebox{\hsize}{!}{\includegraphics{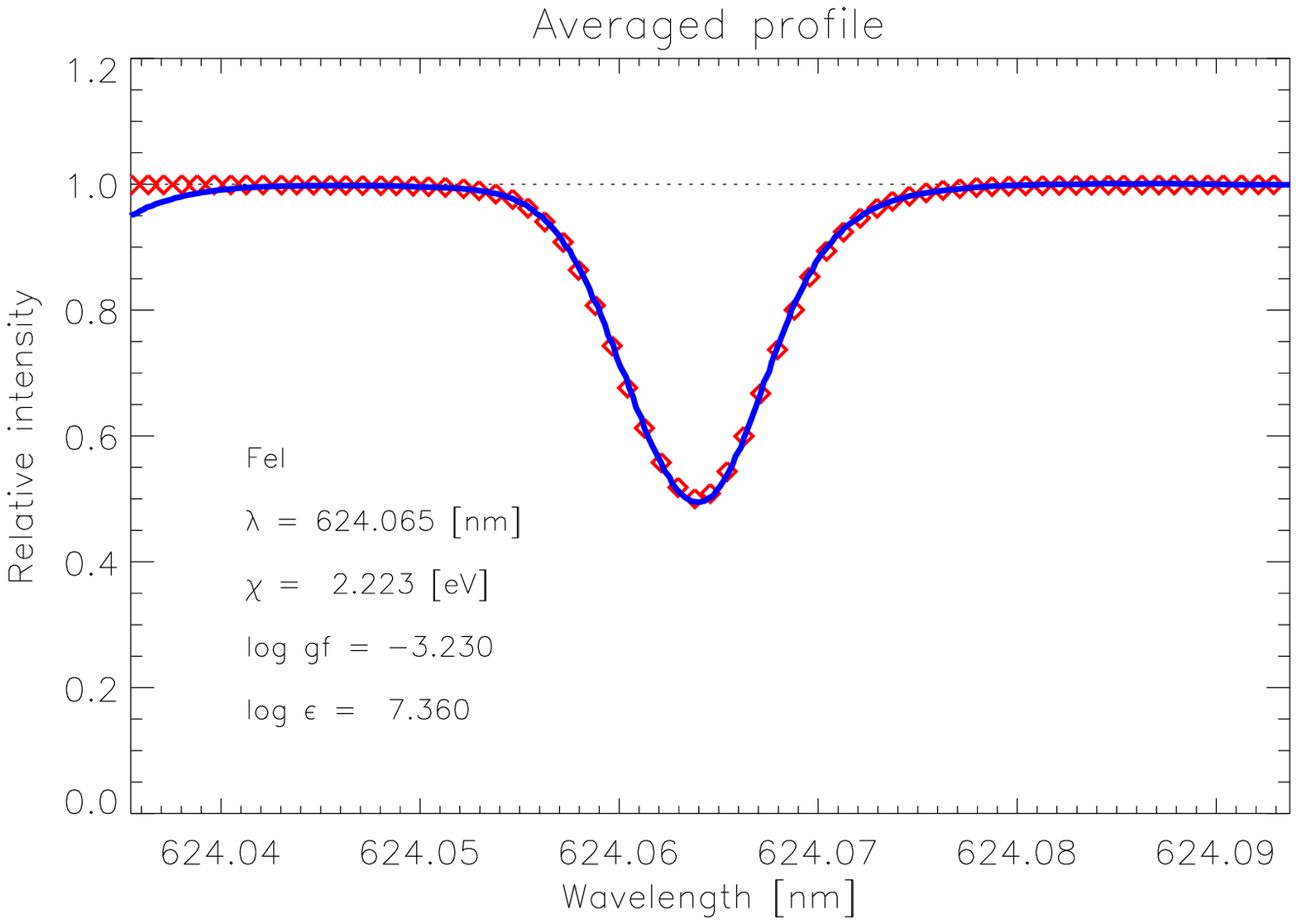}
\includegraphics{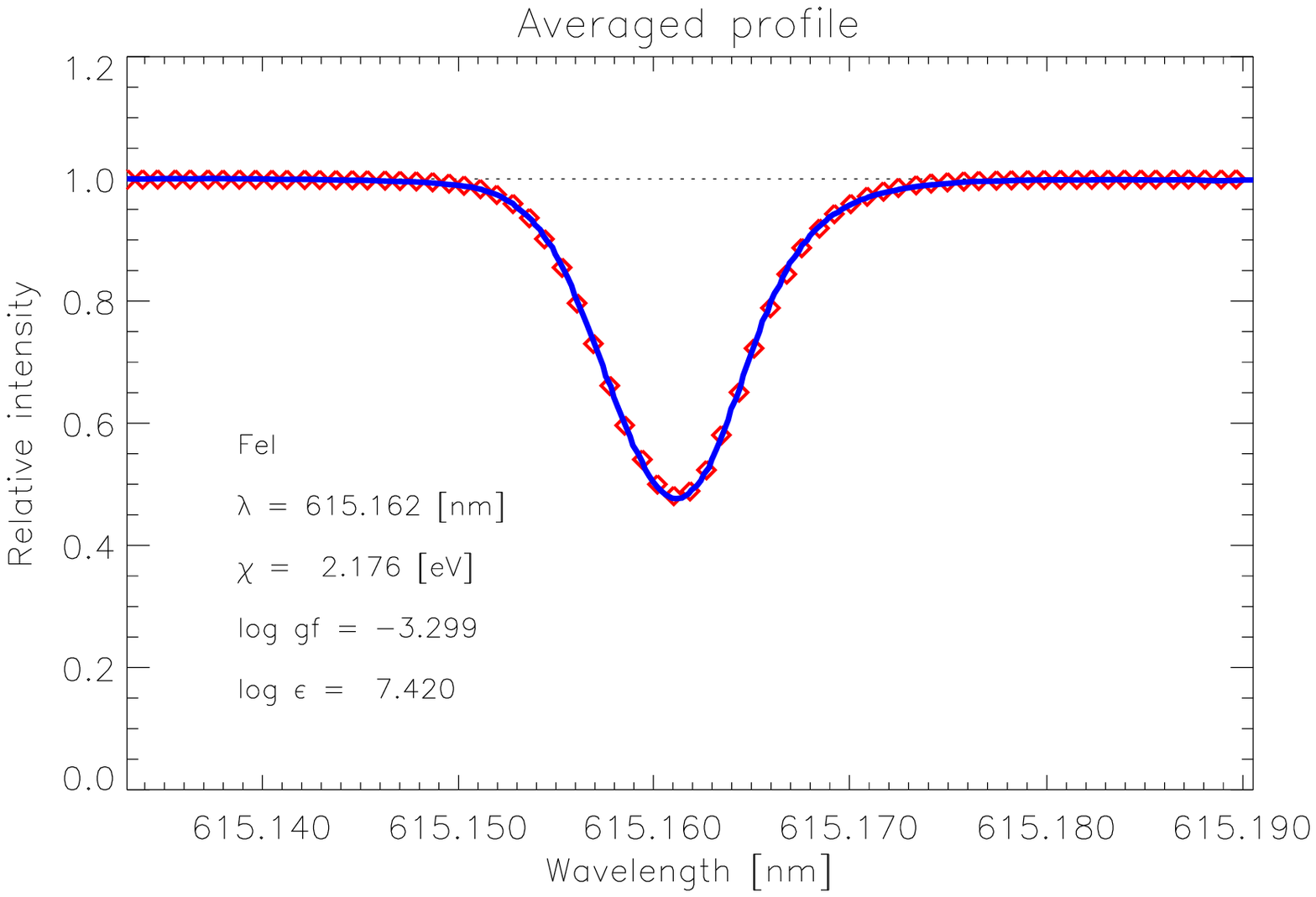}}
\caption{A few comparisons between the predicted (diamonds) 
and observed (solid lines) spatially and temporally averaged 
Fe\,{\sc i} lines at disk-center ($\mu = 1.0$). Only every other point
in the theoretical profiles are shown for clarity. To illustrate the
vast improvement over classical 1D model atmospheres, in the case 
of the Fe\,{\sc i} 615.2\,nm line (lower right panel)
the corresponding prediction with
the Holweger-M\"uller (1974) model atmosphere (dashed line) when adopting a 
microturbulence of 0.845\,km\,s$^{-1}$
and a Gaussian macroturbulence of 2.4\,km (the radial-tangential 
macroturbulence broadening is of course not applicable for intensity 
spectra) is also shown. 
The Fe abundance for the 1D profile has been adjusted to 
return the same equivalent width as the 3D profile 
(${\rm log}\, \epsilon_{\rm Fe I} = 7.59$) and the macroturbulence was
determined by having the same line depths in 1D and 3D. The lack of
line shift and asymmetry for the theoretical 1D profile is clearly seen.
Note that all profiles are shown on an absolute wavelength scale with
no arbitrary wavelength shifts
}
         \label{f:feiprof}
\end{figure*}

\begin{figure}[t!]
\resizebox{\hsize}{!}{\includegraphics{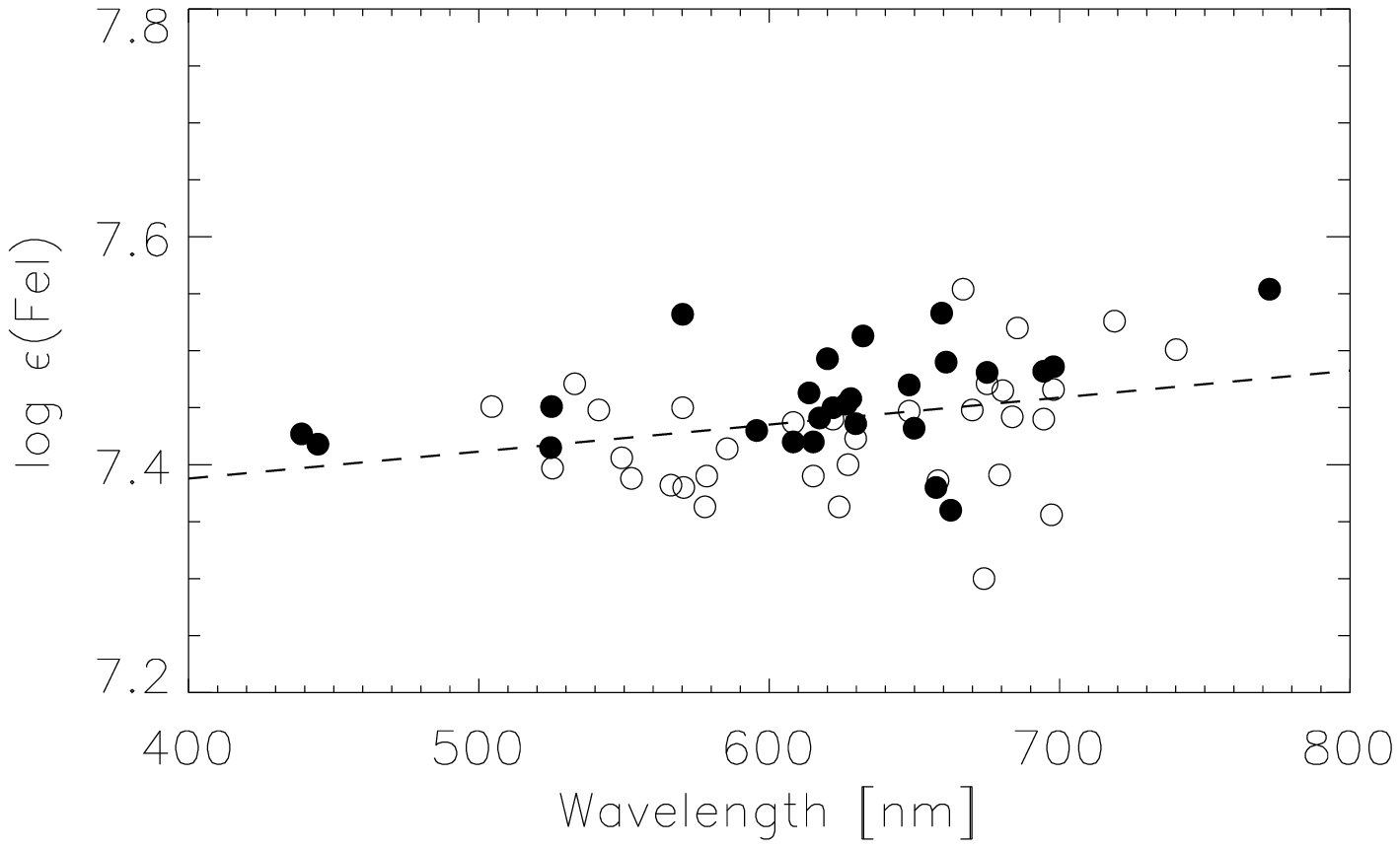}}
\resizebox{\hsize}{!}{\includegraphics{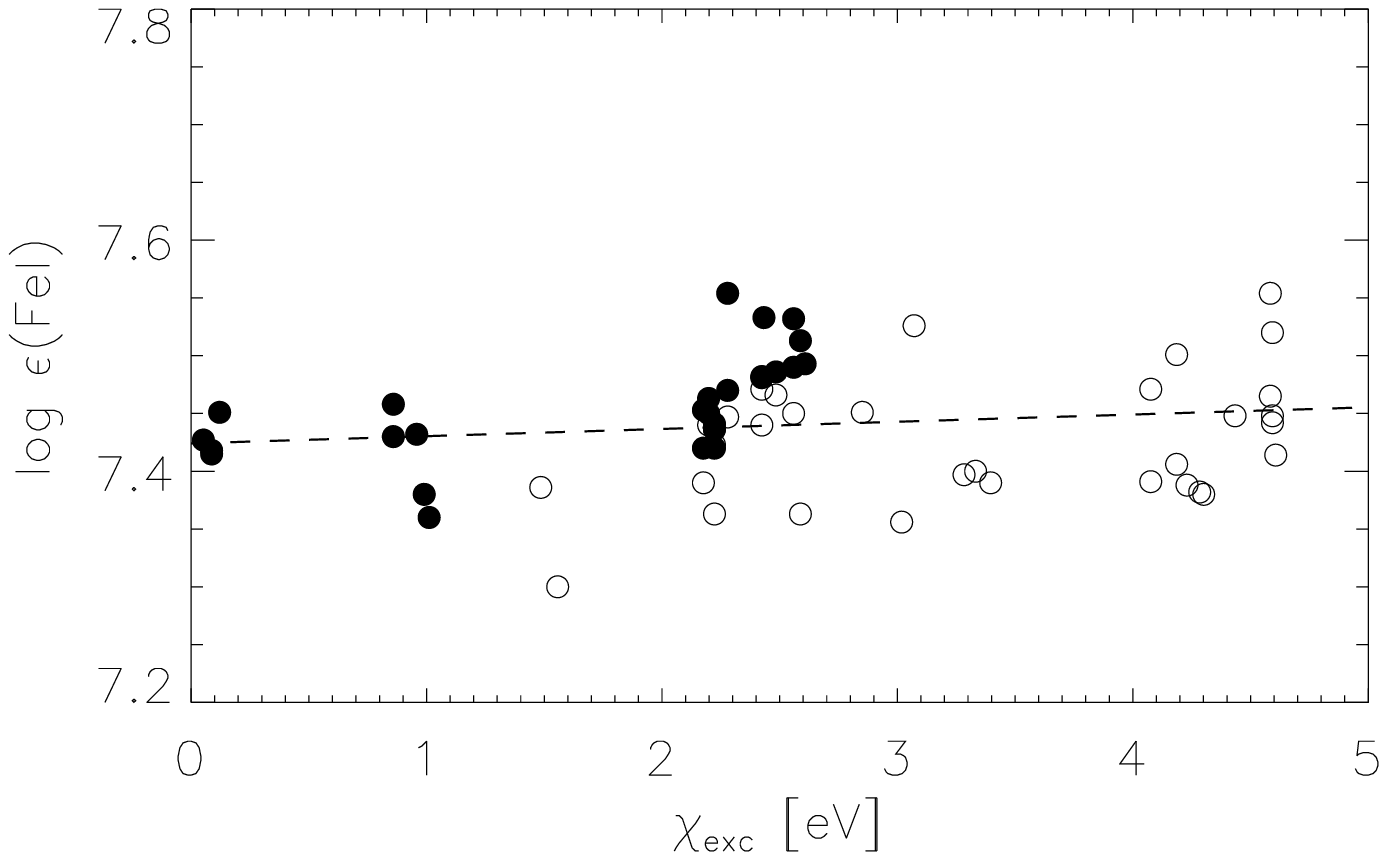}}
\resizebox{\hsize}{!}{\includegraphics{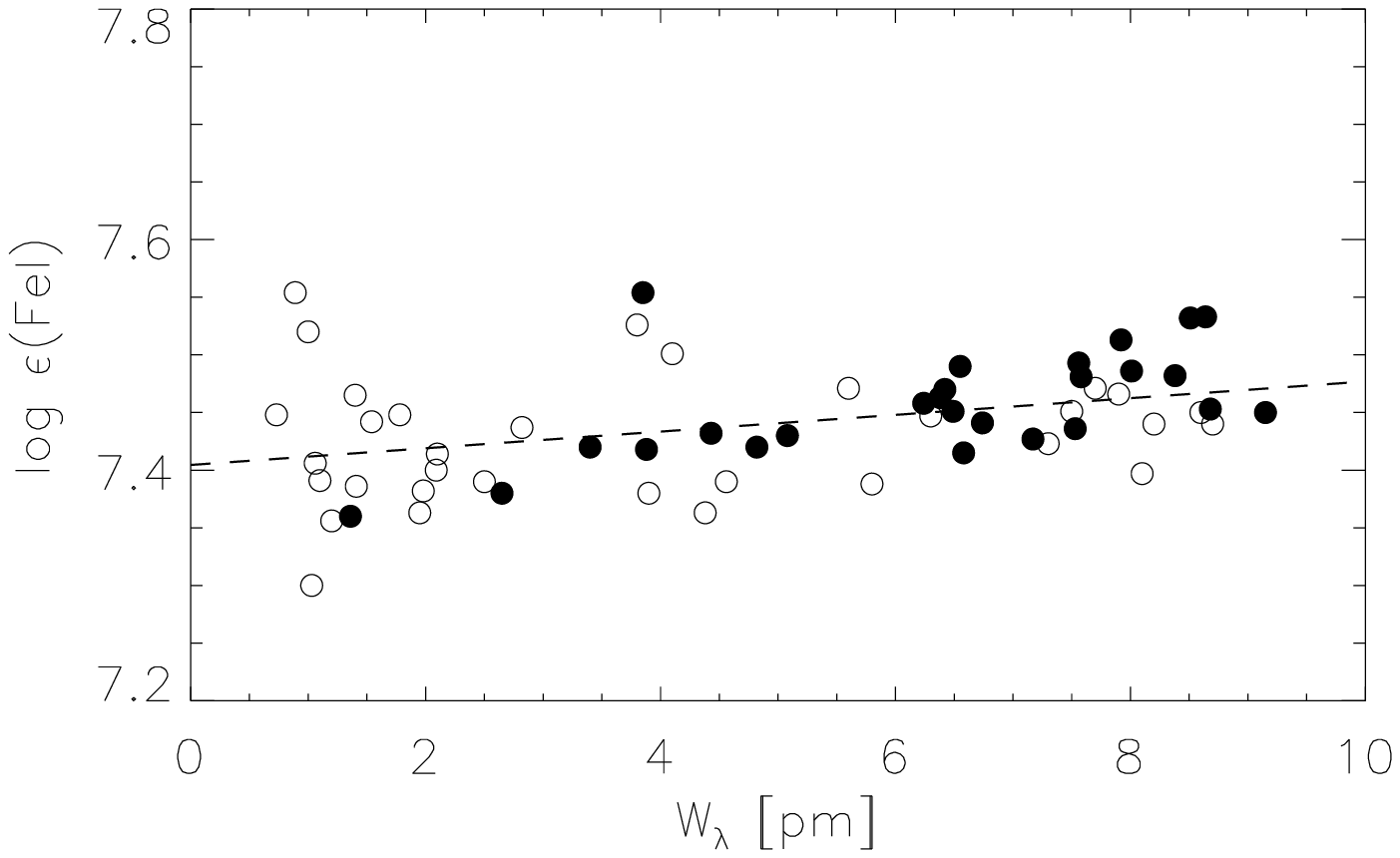}}
\caption{The derived Fe abundance from weak and intermediate-strong
Fe\,{\sc i} lines as functions of  wavelength ({\it Upper panel}),
excitation energy ({\it Middle panel}) of the lower level, and
line strength ({\it Lower panel}). The lines from the Blackwell 
et al. (1995a) and Holweger et al. (1995) samples are marked 
with $\bullet$ and $\circ$, respectively. The dashed lines
are linear least square fits to the data when including all lines
}
         \label{f:fei}
\end{figure}

The derived Fe abundances obtained from profile fitting of the observed
weak and intermediate strong Fe\,{\sc i} lines
are listed in Table \ref{t:fei}. 
We emphasize that the abundances 
have been derived without invoking any
equivalent widths, microturbulence or macroturbulence, leaving the
elemental abundance as the only free parameter which is determined
by the line strength. The good agreement between predicted and observed
profiles is illustrated in Fig. \ref{f:feiprof}; additional examples
can be found in Paper I. It is interesting to contrast the remarkable
consonance achieved with 3D hydrodynamical model atmospheres with 
the results from classical 1D model atmospheres, with or without
macroturbulence (cf. Fig. \ref{f:fei}; 
Paper I; Anstee et al. 1997); the improvement is equally obvious 
and telling.
 
As shown in Fig. \ref{f:fei}, the
individual abundances show no significant dependence on wavelength
or excitation potential. We note that the claimed anomalous lines 
with excitation potential of 2.2\,eV
(Blackwell et al. 1995a) no longer exist in our calculations.
Furthermore, there is no need to fine-tune the temperature structure
to remove existing trends with excitation potential as necessary with the
Holweger-M\"uller (1974) model with which there is about 0.15\,dex 
difference in abundances between low- and high-excitation Fe\,{\sc i}
lines (Grevesse \& Sauval 1999).
There is, however, a slight trend with line
strength in the sense that the strongest lines ($W_\lambda \simeq 9$\,pm)
imply about 0.06\,dex higher abundances than the weakest lines.
The reason for this behaviour will be discussed further in Sect. \ref{s:disc},
although it has a minor impact on the final abundance estimates,
as illustrated below.
It is noteworthy that the trend is more pronounced for the Blackwell
et al. (1995a) sample than for the lines of Holweger et al. (1995),
which may partly explain why different microturbulences were adopted
in the two studies. In terms of a 1D analysis,
the trend would correspond to an underestimated microturbulence of about
0.15\,km\,s$^{-1}$, which emphasizes the relatively minor magnitude of this
shortcoming.
 
The resulting (unweighted) mean abundances of the Oxford and
Hannover-Kiel samples of weak and intermediate strong Fe\,{\sc i} lines are
${\rm log}\, \epsilon_{\rm Fe I} = 7.46 \pm 0.04$ and 
${\rm log}\, \epsilon_{\rm Fe I} = 7.43 \pm 0.05$, respectively,
where the quoted uncertainty is the standard deviation
(twice the standard deviation of the mean = 0.02). 
The difference between the two samples 
essentially reflects the 0.03\,dex offset in the absolute scales. 
From the combined sample the estimate is
${\rm log}\, \epsilon_{\rm Fe I} = 7.44 \pm 0.05$; it should be
noted here that nine lines in common have entered twice into this 
result. For reasons outlined above, we consider the 
transition probabilities of Blackwell et al. (1995a) to be slightly
superior. However, due to the slight trend with line strength our
final (unweighted) Fe determination is still the mean of all lines:
$${\rm log}\, \epsilon_{\rm Fe I} = 7.44 \pm 0.04.$$ 
It should be noted that 
due to the inclusion of two different scales for the
oscillator strengths, the scatter is slightly increased.
The final uncertainty is most likely dominated by systematic
rather than statistical errors, in particular the transition probabilities.
Furthermore, the neglect of NLTE effects and 
observational complications such as blends and continuum level
placement, may introduce additional abundance errors which are of 
comparable size; unfortunately
astronomy has not yet reached the era with 0.02\,dex accuracy in
absolute abundances, in particular not with classical 1D model atmospheres, 
even if it is occasionally claimed in the literature.

The presence of a trend in derived abundances with line strength
have a minor influence on the mean Fe abundance. Restricting the analysis 
to lines with $W_\lambda \le 5$\,pm decreases the mean abundances with
0.03 and 0.01\,dex for the Oxford and Hannover-Kiel samples, respectively,
while leaving the scatter practically intact. The 
larger sensitivity for the former lines can be traced to
their in general larger line strengths. This is likely also the
reason for the slightly larger difference in mean abundances between
the two compilations than accounted for by the
two $gf$-scales. It should be noted, however,
that the Oxford sample only contains seven lines with $W_\lambda \le 5$\,pm,
which may skew the results somewhat; additional weak Fe\,{\sc i} lines
with high-precision furnace oscillator strengths similar to the
published Oxford data would certainly be of great value.

Given the excellent agreement between the predicted line shapes and
observed profiles illustrated in Fig. \ref{f:fei} and Paper I, 
very similar abundances to those presented in
Table \ref{t:fei} would be derived if equivalent widths or
line depths had been used instead of profile fitting. Due to the
larger uncertainties introduced by the subjectivity of equivalent
width measurements and departures from LTE in the line cores, such
abundance diagnostics are significantly more inferior compared
to profile fitting,
provided of course that the model atmosphere is sufficiently realistic
to accurately predict the line profiles (Asplund et al. 2000a). It is 
interesting to note though that adopting the published equivalent widths
of Blackwell et al. (1995) would result in a 0.02\,dex 
{\em higher} Fe\,{\sc i} abundance
while using the  Holweger et al. (1995) values would 
result in a 0.03\,dex {\em lower} abundance than those derived from profile
fitting for the two samples of lines. 

It should be borne in mind that the analysis presented here
assumes LTE, whose validity may be questioned in particular
for Fe\,{\sc i} lines.
Unfortunately no detailed 3D NLTE calculations exist for solar 
Fe lines, and it is therefore difficult to predict how the abundances
in Table \ref{t:fei} would be altered if departures from LTE would
be allowed. Some preliminary guidance may come from 1D NLTE calculations 
(e.g. Solanki \& Steenbock 1988).
The calculations by Shchukina (2000, private communication)
predict an over-ionization of Fe\,{\sc i} and thus that the derived
1D LTE abundances are slightly underestimated by $\la 0.1$\,dex
with the Holweger-M\"uller (1974) model. Simply
adopting these 1D corrections to our 3D LTE results would result in
${\rm log}\, \epsilon_{\rm Fe I} = 7.50 \pm 0.07$ for the combined sample
of Oxford and Kiel lines. Furthermore, the trend with line strength would
become more pronounced ($\simeq 0.13$\,dex difference between
the strongest and weakest lines). Additionally a minor trend with
excitation potential ($\simeq 0.03$\,dex difference between
$\chi_{\rm exc}=0$ and 4.5\,eV transitions with low-excitation lines
returning higher abundances) would appear. However,
we are very reluctant to adopt these results here since it is 
premature to extrapolate 1D predictions to the 3D case until
3D NLTE calculations for Fe exist. 
Because the departures from LTE depend sensitively on the 
adopted model atmosphere, the temperature inhomogeneities may both
amplify or attenuate the 1D NLTE effects. 
Naturally such 3D calculations would be of great interest.

\section{Abundance from strong Fe\,{\sc i} lines
\label{s:feistrong}}

Strong lines have since long been considered less than ideal
for the purposes of abundance determinations due to the poorly
understood collisional broadening which normally requires additional
enhancement factors over the classical Uns\"old (1955) recipe.
Recent progress in the quantum mechanical treatment of the broadening
(Anstee \& O'Mara 1991, 1995; 
Barklem \& O'Mara 1997; Barklem et al. 1998) has, however, opened up
the possibility to use the damping wings of strong lines, which are
little sensitive to the non-thermal broadening affecting
weaker lines, as a complement
to analyses of weaker lines when deriving 
elemental abundances. Anstee et al. (1997) found an excellent agreement
with the meteoritic abundance
for the thus determined Fe abundance from strong Fe\,{\sc i} lines 
and the Holweger-M\"uller (1974) model.

\begin{figure}[t!]
\resizebox{\hsize}{!}{\includegraphics{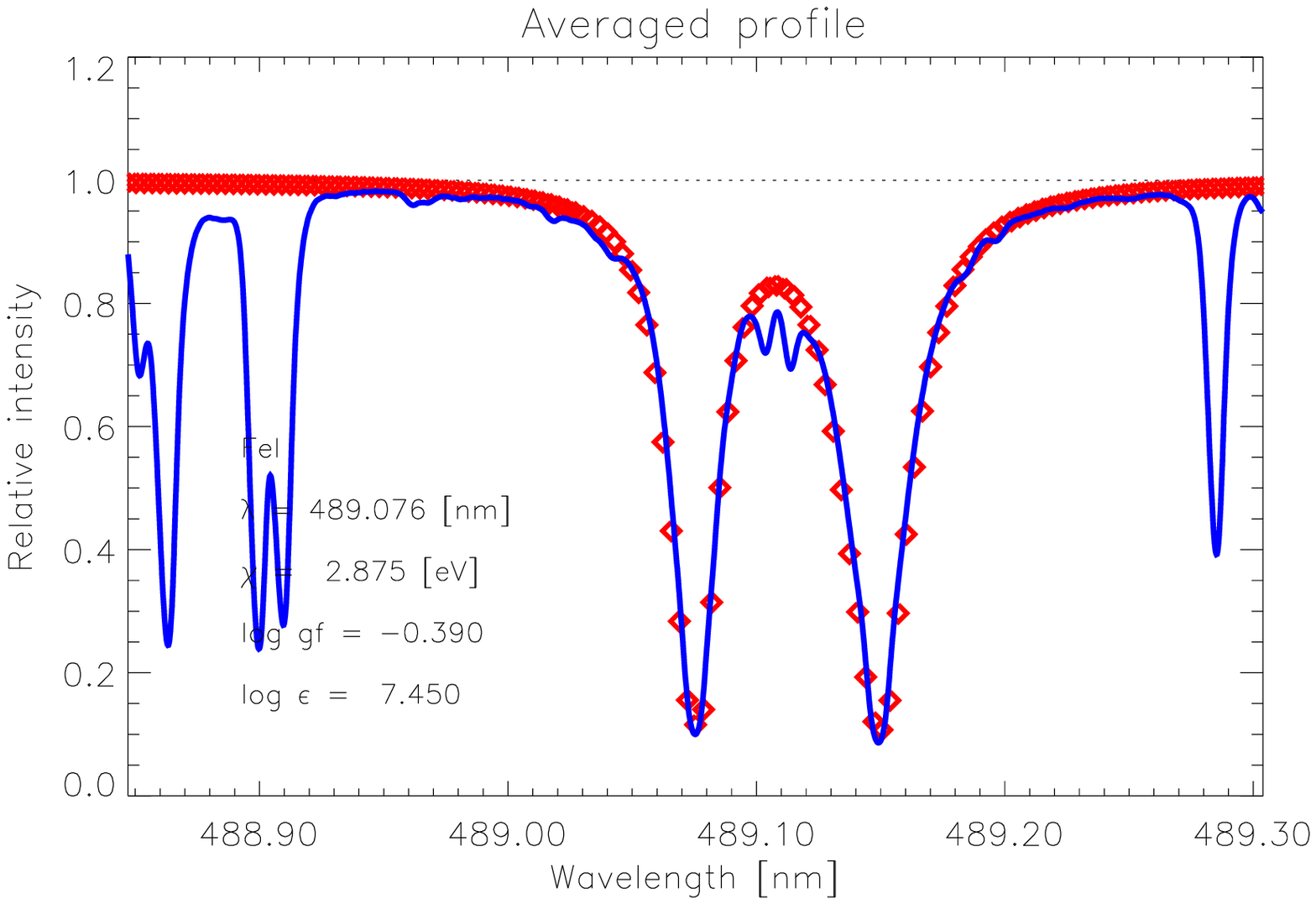}}
\resizebox{\hsize}{!}{\includegraphics{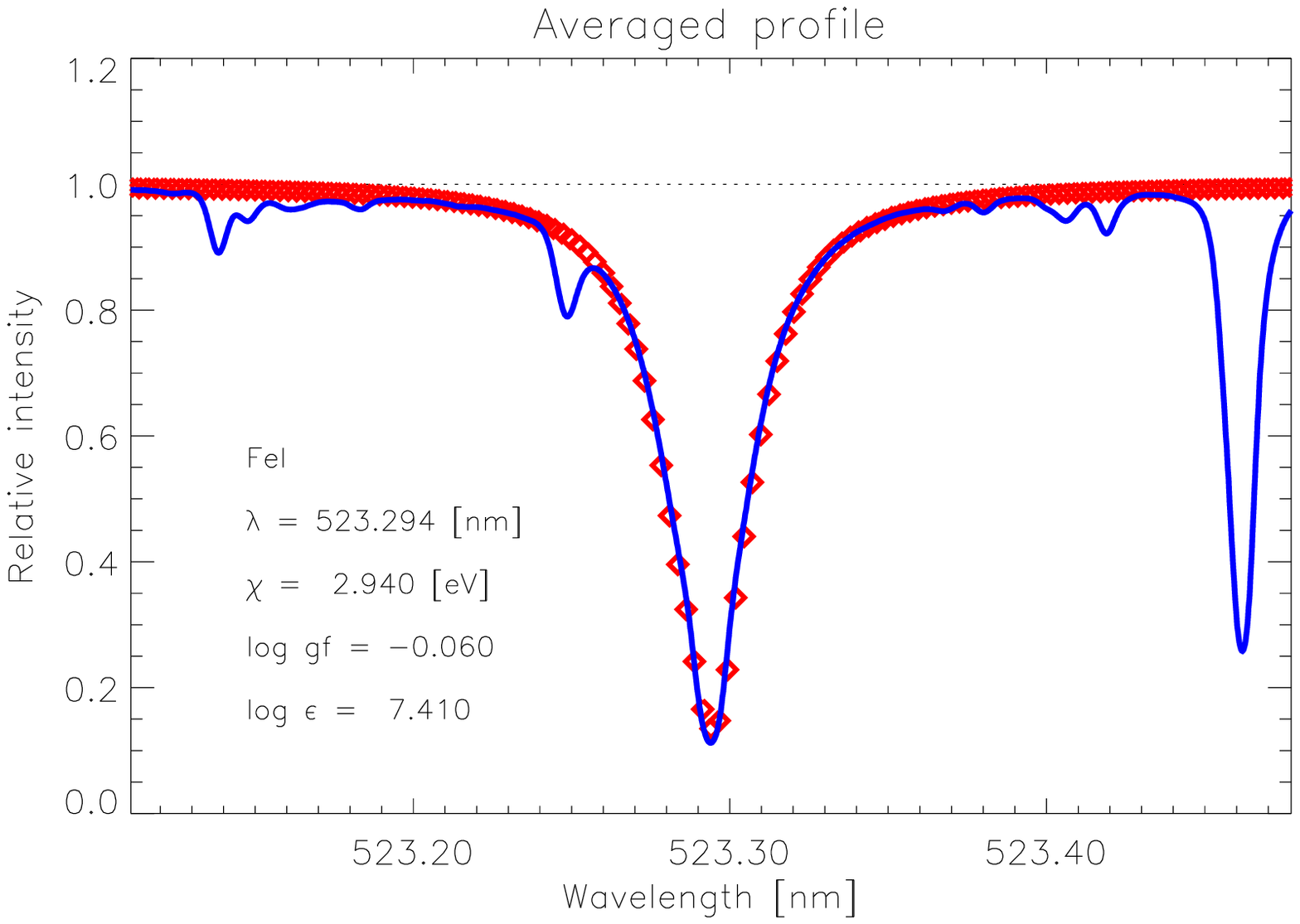}}
\caption{A few examples of the predicted (diamonds) and observed 
(solid lines) strong Fe\,{\sc i} 
lines: Fe\,{\sc i} 489.07+489.15 ({\it Upper panel}) and 523.9\,nm
({\it Lower panel}) lines. Blending lines other than Fe lines have
not been included in the synthesis}
         \label{f:feistrong}
\end{figure}

Table \ref{t:feistrong} lists the derived Fe abundances with 
our 3D hydrodynamical solar atmosphere model using a sample
of strong Fe\,{\sc i} lines which have been considered the most
suitable by Anstee et al. (1997) (lines denoted by quality
category A+, A and A- in their Table 1).
Examples of the obtained agreement between predicted and observed
profiles are found in Fig. \ref{f:feistrong}.
The resulting (weighted) 
mean Fe abundance from the 14 strong Fe\,{\sc i} lines is 
$${\rm log}\, \epsilon_{\rm FeI} = 7.42 \pm 0.03.$$
However, in spite of being considered as very accurate abundance diagnostics
by Anstee et al. (1997),
several of the lines turned out to unsuitable due to uncertainties
introduced by severe blending, continuum placement, radiative broadening
and poorly developed damping wings; those lines are marked in 
Table \ref{t:feistrong} and given half weight in the final abundance
determination. 

Even if the scatter is small for the sample of strong lines, it 
is noteworthy that the standard deviation ($\sigma = 0.03$)
is significantly larger
than the claimed accuracy ($\sigma = 0.01$) 
of the analysis by Anstee et al. (1997) using
the Holweger-M\"uller (1974) model. 
In order to better understand the differences,
we have therefore re-derived Fe abundances for the same lines in an
identical procedure to that of  Anstee et al. (1997), in particular
using their collisional broadening data which differ slightly from
those adopted in Table \ref{t:feistrong} 
which have been provided by Barklem (1999,
private communication) from line-by-line calculations.
Our results with the Holweger-M\"uller (1974) model have a significantly larger
scatter than that quoted by Anstee et al. (1997), identical to
our full 3D analysis of the same lines. We therefore suspect that
the claimed uncertainty in Anstee et al. (1997) is over-optimistic
and that the true scatter is larger, which is also
verified by independent calculations by Barklem (1999, private
communication). The choice of solar atlas (we adopt the more recent
Brault \& Neckel FTS-atlas while Anstee et al. use the older Liege atlas)
has a minor influence on the resulting scatter, although strong lines are
often conspicously asymmetric in the Liege-atlas (e.g. H$\alpha$),
presumably due to inaccurate continuum tracement.
Furthermore there is a systematic offset in theoretical line strengths 
which amounts to about 0.03\,dex in abundance
between our calculations and the identical ones
by Anstee et al. (J. O'Mara, 1999, private communication). 
The reason for this discrepancy is likely due to slight differences
in adopted continuum opacities, the $P_{\rm e}-P_{\rm gas}$-relation 
in the Holweger-M\"uller (1974) model and code implementation.   
This emphasizes again that derived abundances rarely have systematic errors
smaller than 0.02\,dex.

Due to the subjectivity involved with strong lines in terms of
choice of solar atlas,
continuum placement, wavelength shifts, blends, exactly which part
of the wings are given the greatest weight, and remaining uncertainties
in the collisional broadening, we consider abundances 
derived from strong lines to be inferior to those from weaker lines,
although they naturally serve as important complements. In this respect it is
reassuring that the here derived Fe abundance from strong Fe\,{\sc i}
lines agree well with those from weak and intermediate strong
Fe\,{\sc i} and Fe\,{\sc ii} lines presented in Sects. \ref{s:fei} and
\ref{s:feii}.

\section{Abundance from Fe\,{\sc ii} lines
\label{s:feii}}

\begin{figure*}[t!]
\resizebox{\hsize}{!}{\includegraphics{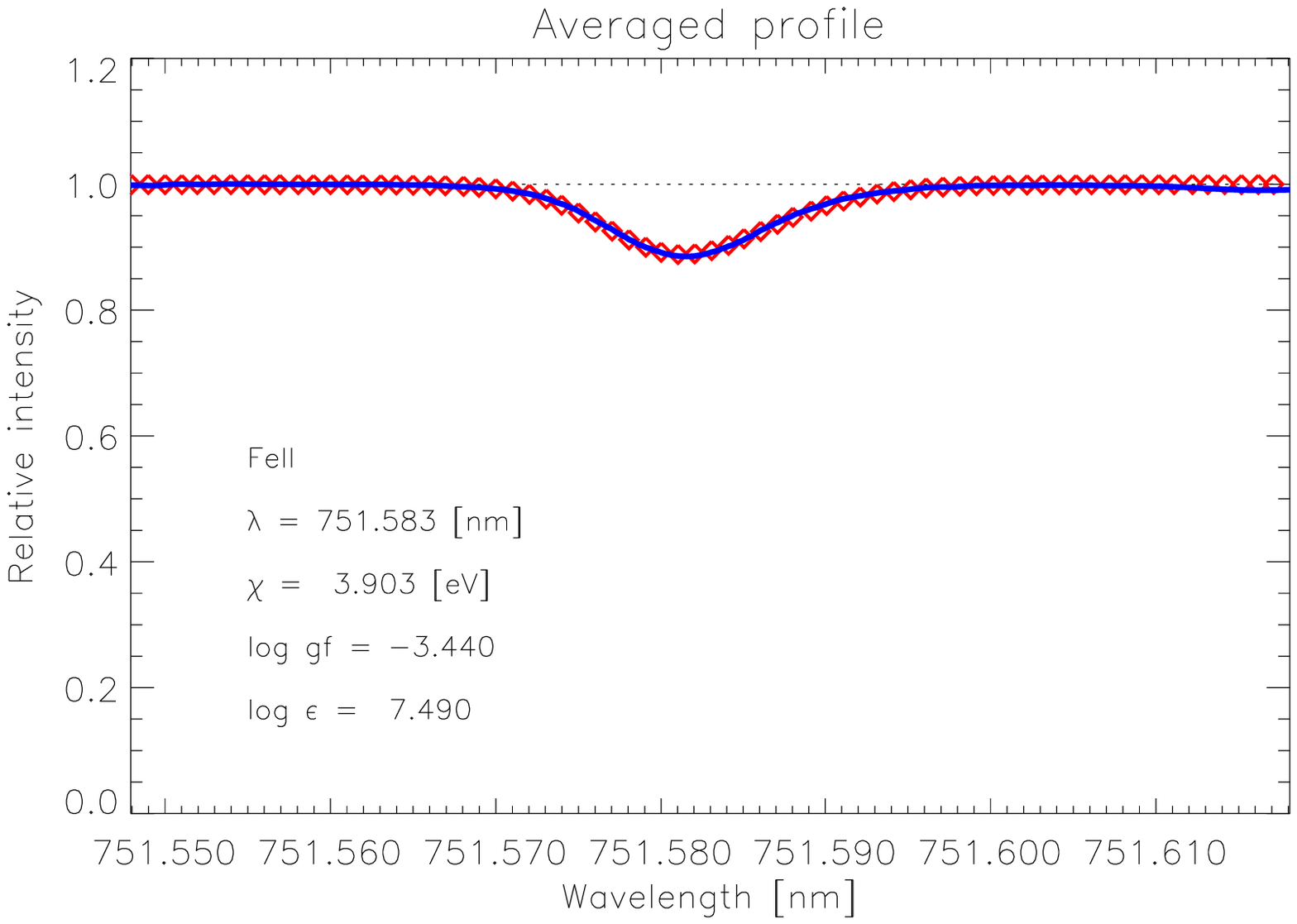}
\includegraphics{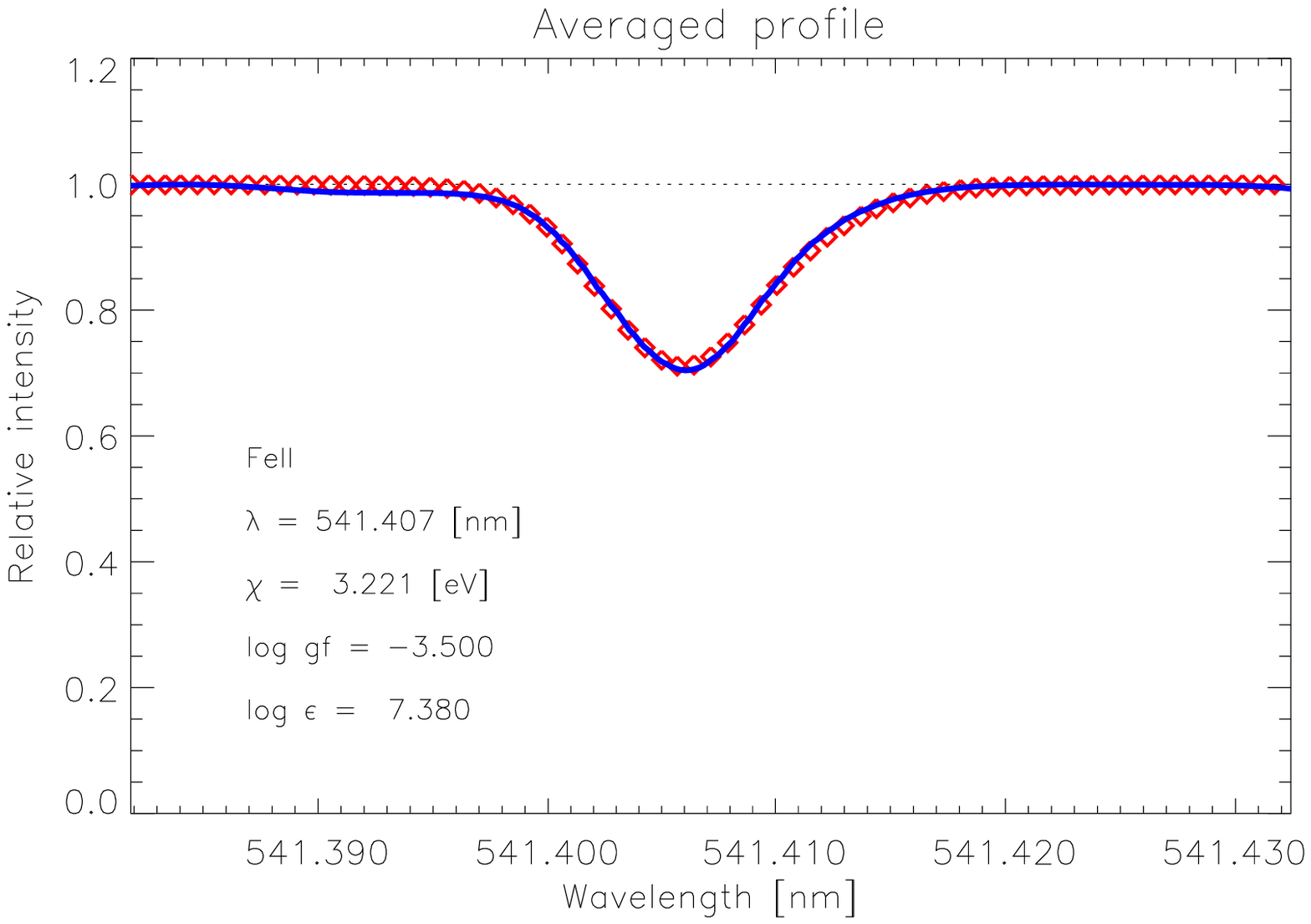}}
\resizebox{\hsize}{!}{\includegraphics{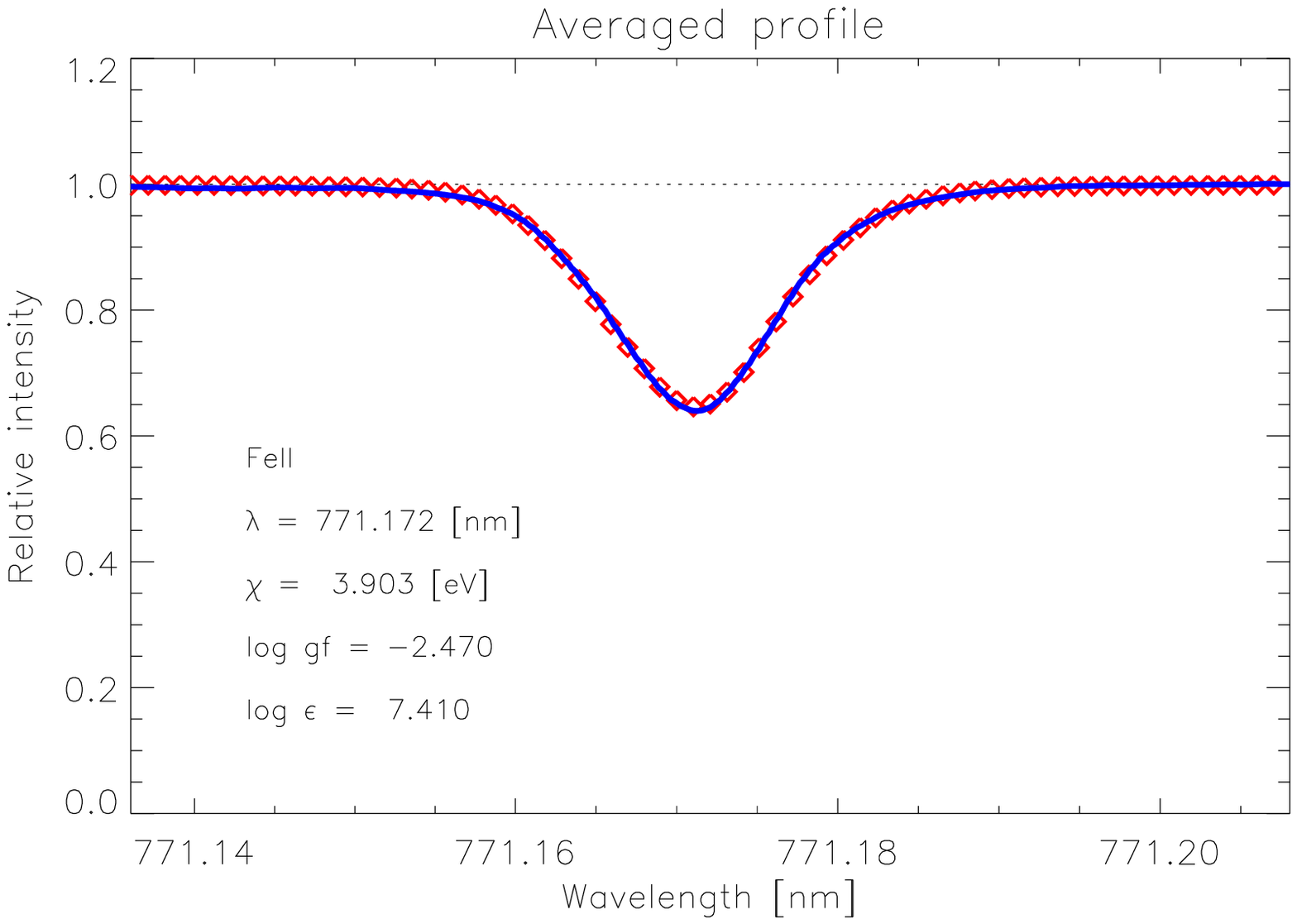}
\includegraphics{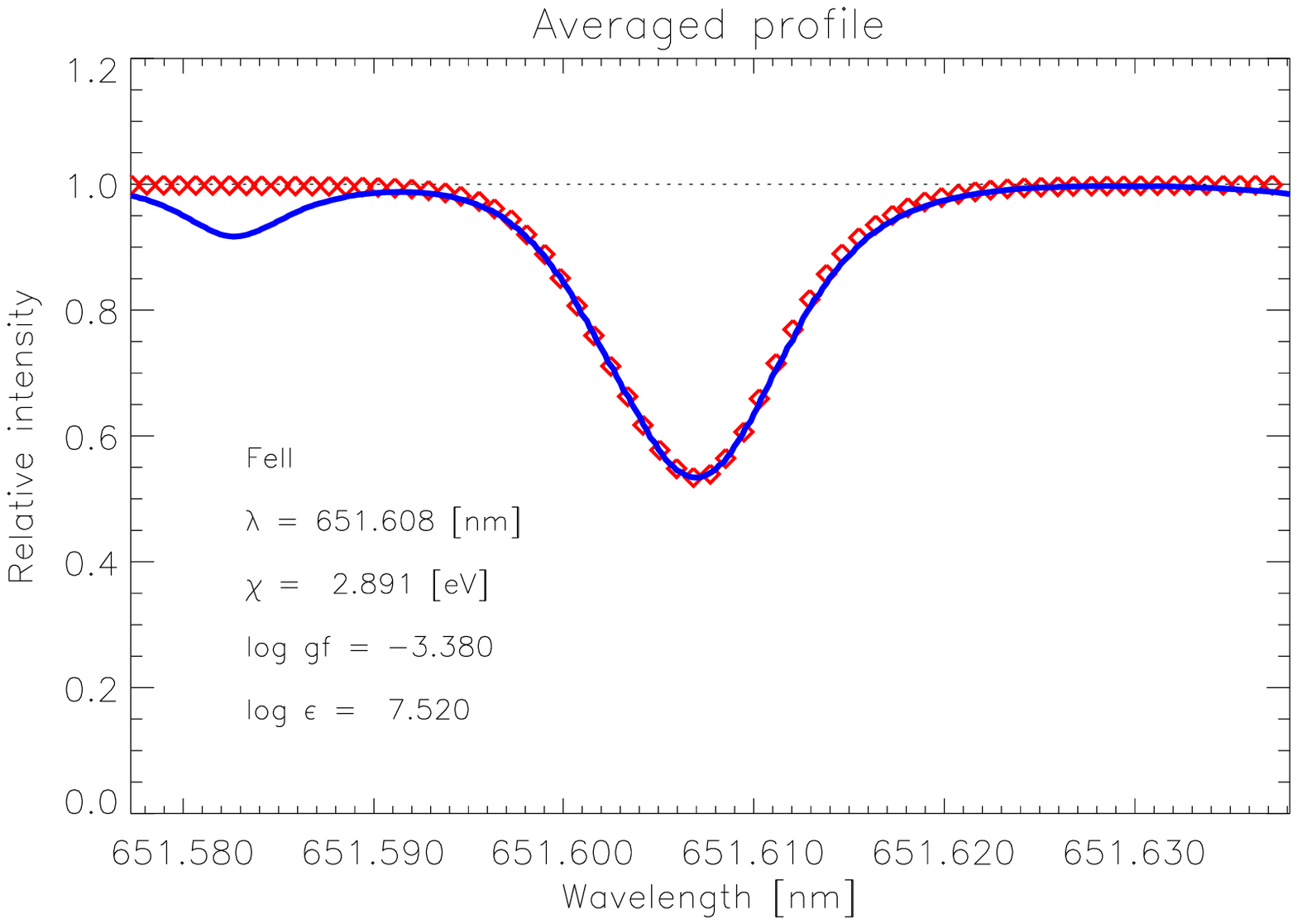}}
\caption{A few comparisons between the predicted (diamonds) 
and observed (solid lines) spatially and temporally averaged 
Fe\,{\sc ii} lines at disk-center ($\mu = 1.0$)
}
         \label{f:feiiprof}
\end{figure*}

The Fe abundance values obtained from the individual Fe\,{\sc ii}
lines are listed in Table \ref{t:feii}; a few examples of the
achieved agreement between theory and observations are given
in Fig. \ref{f:feiiprof}. 
As shown in Fig. \ref{f:feii}, the individual abundances show no
significant dependence on neither the wavelength, excitation 
potential of the lower level (though the adopted lines provide only a 
restricted range), nor the line strength. 
In this respect, the Fe\,{\sc ii} lines differ from the
Fe\,{\sc i} lines, which show a minor trend with the line strength,
at least within the assumption of LTE. 

The resulting (unweighted) mean abundance becomes
$${\rm log}\, \epsilon_{\rm FeII} = 7.45 \pm 0.10,$$
where the quoted uncertainty is the standard deviation
(twice the standard deviation  of the mean = 0.05). 
The quoted error of course only reflects the internal accuracy
and thereby possible uncertainties in e.g. the absolute scale 
of the $gf$-values are not accounted for. 
With Holweger et al.'s $gf$-values the mean abundance would be 0.04\,dex
higher while it would be 0.02\,dex lower with the measurements given 
in Schnabel et al. (1999); 
the estimated error would remain basically unaltered by such exercises. 
As previously noted, with an Uns\"old enhancement factor of 1.5
instead of 2.0 the abundance would be 0.01\,dex higher. When excluding
the discrepant Fe\,{\sc ii}\,744.93\,nm line, which is significantly 
blended in the red wing, the mean abundance is increased by 0.01\,dex;
we also note that the $gf$-value for this line has a comparatively
large uncertainty (Hannaford et al. 1992).
Restricting to the ten lines with $W_\lambda < 5.0$\,pm, increases the
mean abundance by only 0.006\,dex, demonstrating that the observed trend with
line strength only affects the Fe\,{\sc i} lines. 

The main advantage with Fe\,{\sc ii} lines is their low
sensitivity to details of the temperature structures and
departures from LTE due to over-ionization.
Furthermore, their weakness
ensures that the lines are formed in the
deeper layers which are less susceptible to NLTE excitation effects
such as photon pumping and suction.
The abundance derived from Fe\,{\sc ii} lines should therefore
be an accurate measure of the solar Fe abundance, provided the
transition probabilities are reliable enough. It is
reassuring that our average is in good agreement with the meteoritic
value $7.50 \pm 0.01$ (Grevesse \& Sauval 1998),
in particular in view of the uncertainty in the absolute 
$gf$-scales (Holweger et al. 1990; 
Hannaford et al. 1992; Schnabel et al. 1999) and that the meteoritic
abundance scale probably needs to be adjusted downward by about 0.04\,dex
due to the revised photospheric Si abundance (Paper III).

\section{Discussion
\label{s:disc}}

The results presented in Sects. \ref{s:fei}, \ref{s:feistrong}
and \ref{s:feii} paint a consistent picture for the
solar photospheric Fe abundance. Both the weak Fe\,{\sc i} and
Fe\,{\sc ii} lines suggest very similar abundances:
${\rm log}\, \epsilon_{\rm FeI} = 7.44\pm0.05$ and 
${\rm log}\, \epsilon_{\rm Fe} = 7.45\pm0.10$.
Since this result does not rely on equivalent widths, microturbulence,
macroturbulence, or, at least for the Fe\,{\sc i} lines, collisional
damping enhancement factors, and is based on highly realistic 3D,
hydrodynamical model atmospheres, it seems like the long-standing
solar Fe problem (e.g. Blackwell et al. 1995a,b; Holweger et al. 1995)
has finally been settled in favour of the meteoritic value, 
in particular considering the slight revision recently of the photospheric
Si abundance and thus the whole absolute scale for the meteoritic
abundances (Paper III): ${\rm log}\, \epsilon_{\rm Fe} = 7.46 \pm 0.01$.
In fact the agreement between the photospheric and meteoritic values
is partly fortuitious since the remaining uncertainties in oscillator
strengths and model atmospheres are likely on the order of 0.04\,dex.

The difference of 0.18\,dex (7.64 vs 7.46) between our result and the one by
Blackwell et al. (1995a) using the same set of $gf$-values is
attributable to a switch to line profile fitting and improved
collisional broadening treatment, and exchange of the microturbulence
concept for self-consistent Doppler broadening from convective motions
and the Holweger-M\"uller (1974) model for an ab initio 3D hydrodynamical
model atmosphere.
Our (unweighted) 
Fe\,{\sc ii} result ${\rm log}\, \epsilon_{\rm FeII} = 7.45 \pm 0.05$ 
is similar to the (weighted) mean
${\rm log}\, \epsilon_{\rm FeII} = 7.47 \pm 0.04,$ found by 
Hannaford et al. (1991) using the same $gf$-values and
the Holweger-M\"uller (1974) semi-empirical model atmosphere, 
which reflects the small
sensitivity of the Fe\,{\sc ii} lines to the details of the
model atmospheres; indeed when instead of profile fitting the equivalent
widths of Hannaford et al. (1991) 
are adopted the (unweighted) mean with the 3D solar model atmosphere
is ${\rm log}\, \epsilon_{\rm FeII} = 7.47 \pm 0.04$.
The main systematic error is
no longer the model atmospheres and analysis as such, but is likely
dominated by the accuracy of the transition probabilities, which still
is on the level of 0.03\,dex on average for both  Fe\,{\sc i} and
Fe\,{\sc ii} lines, even though the {\em internal} precision
may be higher. 

It is of interest to compare our findings for the solar
Fe abundance with previously published
studies based on 2D and 3D hydrodynamical models of the
solar photosphere.
Atroschenko \& Gadun (1994) discuss derived Fe abundances from
Fe\,{\sc i} and Fe\,{\sc ii} lines based on two different types
of 3D model atmospheres (with $30^3$ and $32^3$ gridpoints, respectively,
to compare with our simulation with the dimension 200\,x\,200\,x\,82) 
but obtain significantly more discrepant 
results than those presented here:
${\rm log}\, \epsilon_{\rm Fe I} = 7.05 \pm 0.06$,
${\rm log}\, \epsilon_{\rm Fe II} = 7.48 \pm 0.03$ and
${\rm log}\, \epsilon_{\rm Fe I} = 7.61 \pm 0.02$,
${\rm log}\, \epsilon_{\rm Fe II} = 7.42 \pm 0.02$, respectively;
here the astrophysically determined $gf$-values for the 
Fe\,{\sc ii} lines (using ${\rm log}\, \epsilon_{\rm Fe} = 7.64$)
have been rescaled to agree with the ones by
Hannaford et al. (1992) which we have adopted. These results are,
however, based on equivalent widths for selected lines
and the use of microturbulence,
which had to be introduced in an attempt to hide a very conspicious
trend with equivalent width. Furthermore, the estimated abundances
only made use of the very weakest lines and therefore represent underestimates
for the Fe\,{\sc i} lines. The discrepancies can likely be attributed to
the use of gray opacities for the 3D model atmospheres and too small height extension, resolution and temporal sampling (the
spectral synthesis was restricted to only one respectively two snapshots
from the two simulation sequences and therefore 
should not be considered as proper temporal
averages). These problems with not sufficiently realistic model
atmospheres are also manifested in the relatively poor
agreement with observed line profiles and asymmetries.

\begin{figure}[t!]
\resizebox{\hsize}{!}{\includegraphics{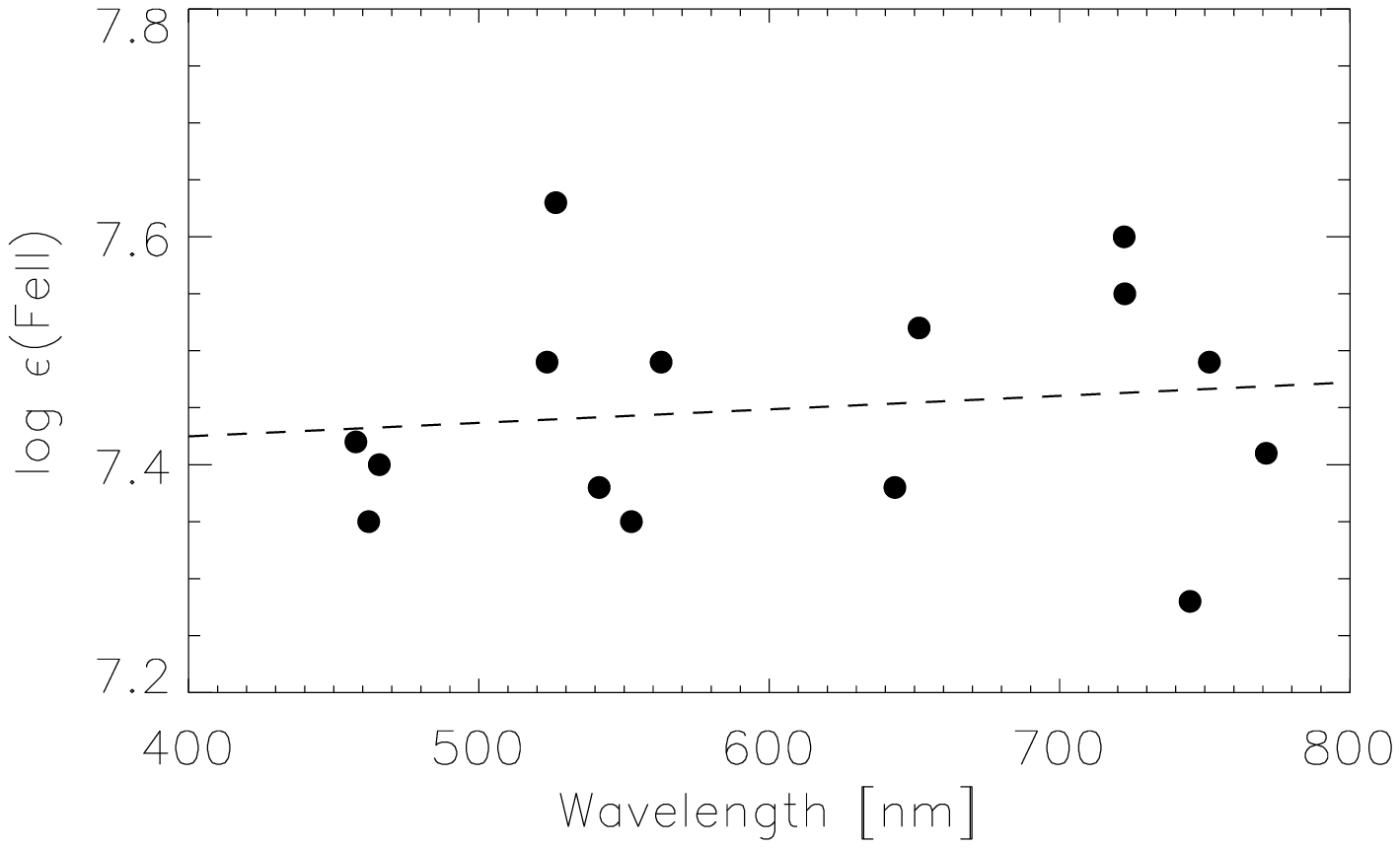}}
\resizebox{\hsize}{!}{\includegraphics{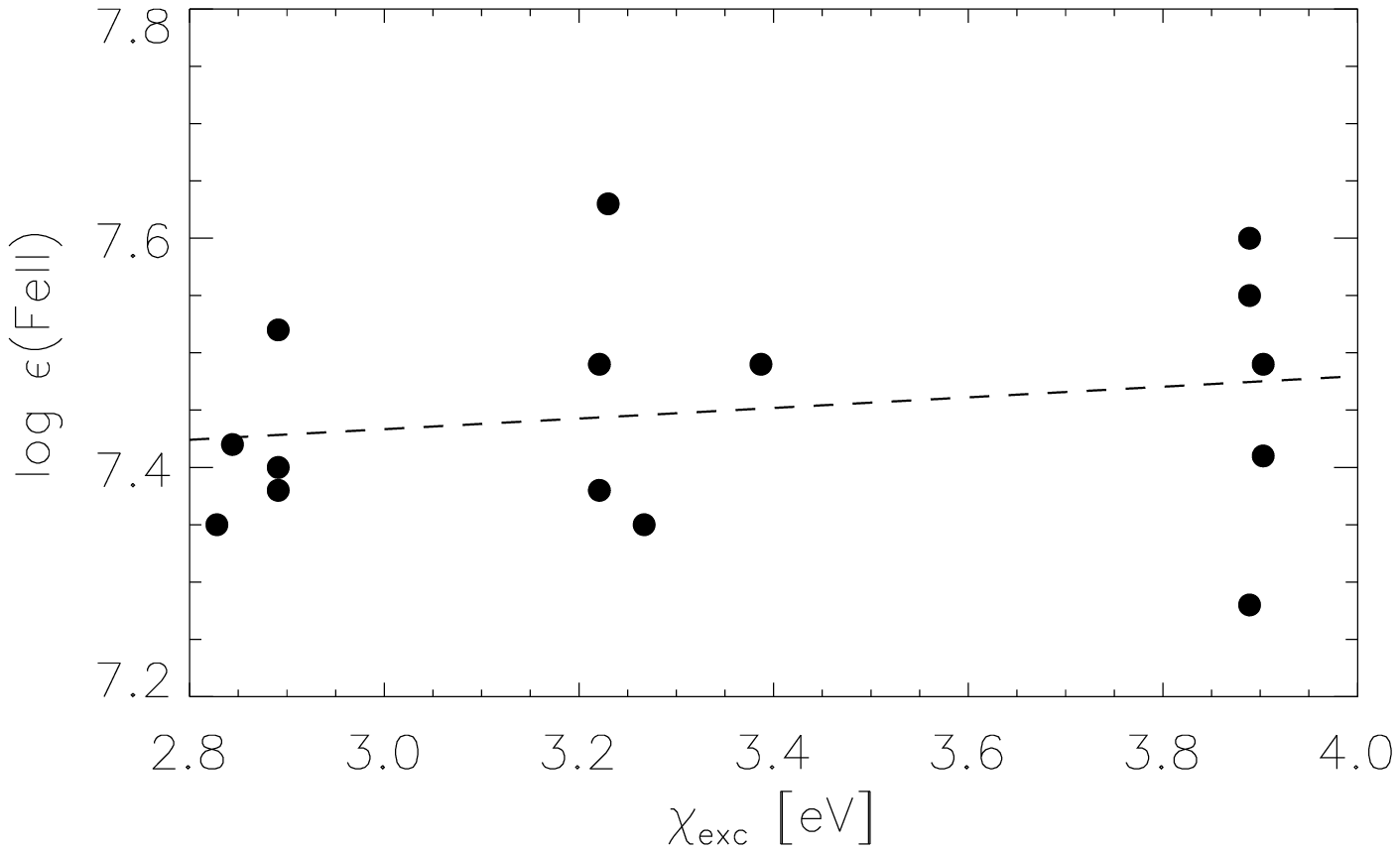}}
\resizebox{\hsize}{!}{\includegraphics{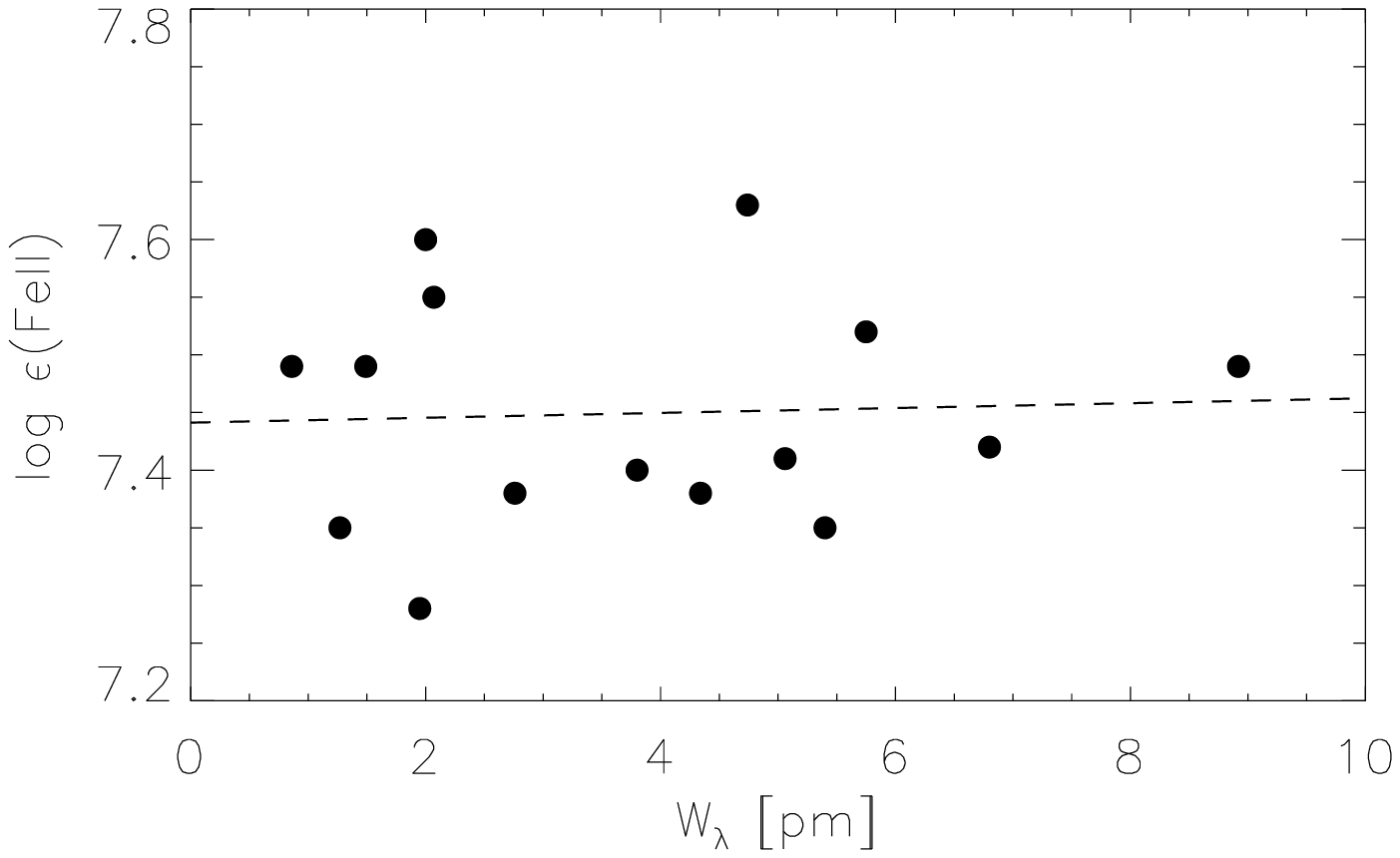}}
\caption{The derived Fe abundance from weak and intermediate-strong
Fe\,{\sc ii} lines as functions of wavelength ({\it Upper panel}),
excitation energy of the lower level ({\it Middle panel}), and
line strength ({\it Lower panel}). The dashed lines
are linear least square fits to the data
}
         \label{f:feii}
\end{figure}

The study of Gadun \& Pavlenko (1997) suffer from similar 
problems. Their model atmospheres were 2D solar convection simulations
(with 112\,x\,58 gridpoints)
but properly temporally averaged. Utilizing equivalent widths, they
derive ${\rm log}\, \epsilon_{\rm Fe I} = 7.33 \pm 0.06$ and
${\rm log}\, \epsilon_{\rm Fe II} = 7.44 \pm 0.02$ for their most reliable
simulation sequence; again the Fe\,{\sc ii} result have been rescaled 
for consistency with our analysis.
Unfortunately, they do not show any comparison between predicted and
observed line profiles, but judging from the
differences in Fe abundances when derived from equivalent widths and
line depths we conclude that the theoretical line profiles 
are too narrow, a common problem with a too poor numerical
resolution in the simulations (Asplund et al. 2000a). 
To summarize, we are confident that our analysis is superior to previously 
published studies with multi-dimensional model atmospheres, a 
conclusion which is further supported by the excellent agreement
between the predicted line profiles and asymmetries with observations,
as described in detail in Paper I, and the confluence between the
Fe\,{\sc i}, Fe\,{\sc ii} and meteoritic results (Paper III).

As noted in Sects. \ref{s:fei} and \ref{s:feii}, neither the
Fe\,{\sc i} nor the Fe\,{\sc ii} results depend on the 
wavelengths or the excitation potentials of the lines.
Furthermore, the Fe\,{\sc ii} lines show no trend with line
strength, in spite of no microturbulence has entered the analysis,
which, as explained in Paper I, is a consequence of the non-thermal
Doppler broadening from the self-consistently calculated 
convective velocity field. However, according to Fig. \ref{f:fei}
the individual Fe\,{\sc i} abundances appear to depend slightly on
the line strength, which could signal an underestimated rms
vertical velocity in the line forming layers of the solar simulation
(Paper I) or a too poor numerical resolution (Asplund et al. 2000a).
But considering the good overall agreement for the theoretical
and observed line shapes, which if anything suggests a slightly
overestimated rms velocity (Paper I) and that no corresponding trend
is present for the Fe\,{\sc ii} (Fig. \ref{f:feii})
and Si\,{\sc i} (Paper III) lines,
this conclusion seems less likely. Instead we suggest the existence
of minor departures from LTE in the stronger lines, which causes
the Fe abundances of these to be slightly overestimated. Such 
departures are more likely to affect Fe\,{\sc i} than
Fe\,{\sc ii} lines and furthermore stronger lines are more
susceptible than weak lines due to the decoupling of the non-local
radiation field and local kinetic gas temperature in the
higher atmospheric layers. Clearly 
an investigation of possible NLTE effects for Fe with 3D inhomogeneous
model atmospheres would be interesting, similarly to the recent
3D calculations for Li (Kiselman 1997, 1998; Asplund \& Carlsson 2000).

\section{Conclusions}

The application of ab initio 3D hydrodynamical model atmospheres
of the solar photosphere to the line formation of Fe\,{\sc i}
and Fe\,{\sc ii} lines has allowed an accurate determination 
of the solar photospheric Fe abundance. Since such a procedure
does not invoke any free adjustable parameters besides 
the treatment of the
numerical viscosity in the construction of the 3D, time-dependent,
inhomogeneous model atmosphere and the elemental abundance in the 3D spectral
synthesis, and considering that whole line profiles
are fitted rather than equivalent widths, the results should provide
a more secure abundance determination than previously accomplished.
The confusion introduced by the various choices of mixing length parameters,
microturbulence and macroturbulence no longer needs to cloud the conclusions. 
Furthermore, the analysis has made use of recent quantum mechanical 
calculations for the collisional broadening of the Fe\,{\sc i}
lines (Anstee \& O'Mara 1991, 1995; Barklem \& O'Mara 1997;
Barklem et al. 1998), which removes the problematical damping
enhancement parameters normally employed, at least for the Fe\,{\sc i}
lines. In view of these
improvements, it is a significant accomplishment that a consistent
picture is emerging in terms of Fe abundances: Fe\,{\sc i}
and Fe\,{\sc ii} lines suggest ${\rm log}\, \epsilon_{\rm Fe} = 7.44\pm0.05$
and ${\rm log}\, \epsilon_{\rm Fe} = 7.45\pm0.10$, respectively,
which agree very well with the meteoritic value 
${\rm log}\, \epsilon_{\rm Fe I} = 7.46 \pm 0.01$ (Paper III)
given the remaining uncertainties in the transition
probabilities. Fe\,{sc i} lines may be slightly more susceptible
for departures from LTE but on the other hand 
the $gf$-values for Fe\,{\sc ii} lines are 
somewhat less accurate. Our final best 
estimate for the photospheric Fe abundance
is therefore simply the average of the two results, until detailed
3D NLTE calculations and improved measurements of the transition 
probabilities have been performed.  
Finally, the debate of the photospheric Fe abundance
(e.g. Blackwell et al. 1995a,b; Holweger et al. 1995)
seems to have been settled in favour of the low
meteoritic abundance. Also strong Fe\,{\sc i} lines imply a similar
photospheric abundance: ${\rm log}\, \epsilon_{\rm FeI} = 7.42 \pm 0.03$
although we give this result lower weight due to the difficulties
involved in analysing the wings of strong lines.

When comparing our results with other recent investigations of
the solar Fe abundance, it is natural to ask why our study should
be preferred. After all, traditional analyses using classical
1D model atmospheres, such as the Holweger-M\"uller (1974) model, has long
been considered sufficient. However, as  
stricter demands are placed on the results in terms of accuracy,
an improved analysis is required. Why should one
embrace the results based on hydrostatic 1D model atmospheres,
equivalent widths and ad-hoc broadening through 
microturbulence and macroturbulence, when such 1D models are inferior 
to the here presented ab initio 3D hydrodynamical models in terms of
the observational diagnostics available for the Sun, such as
granulation topology, velocities and statistics, time-scales and length-scales
of the convection, continuum intensity brightness contrast,
detailed spectral line profiles, asymmetries and shifts, flux distribution,
limb-darkening and H-line profiles (e.g. Stein \& Nordlund 1998;
Asplund et al. 1999b; Paper I)? 
We leave it for the reader to ponder this rhetorical question.

\begin{acknowledgements}
It is a pleasure to thank Paul Barklem, Sveneric Johansson and
Jim O'Mara for helpful discussions and for providing unpublished
line data in terms of line broadening and laboratory wavelengths.
Discussions with Natalia Shchukina regarding departures from
LTE for Fe are much appreciated, as are  
the constructive suggestions by an anonymous referee.
Extensive use have been made of the VALD database
(Kupka et al. 1999), which is gratefully acknowledged.
\end{acknowledgements}


\end{document}